\documentclass{ws-p10x7}

\usepackage{array,euscript}
\usepackage{amsmath}
\usepackage{amssymb}

\newcommand\figsize{0.48\textwidth}

\newcommand\figsizeb{0.37\textwidth}
\newcommand\figsizec{0.4\textwidth}
\def\Journal#1#2#3#4{{#1} {\bf #2}, #3 (#4)}

\def\NPB{{\em Nucl. Phys.} B}
\def\PLB{{\em Phys. Lett.}  B}
\def\PRL{\em Phys. Rev. Lett.}
\def\PRD{{\em Phys. Rev.} D}

\def\PRC{{\em Phys. Rev.} C}
\def\CPC{\em Comp. Phys. Comm.}  
\def\PPNP{\em Prog. Part. Nucl. Phys.} 
\def\NPA{{\em Nucl. Phys.} A} 
\def\EPJC{{\em Eur. Phys. J.} C}

\newcommand     \be             {\begin{equation}}
\newcommand     \ee             {\end{equation}}
\newcommand     \bea             {\begin{eqnarray}}
\newcommand     \eea             {\end{eqnarray}}

\begin{document}

\title{Polarized Structure Functions \\ and \\ Spin Physics}

\author{Uta St\"osslein}

\address{Nuclear Physics Laboratory, University of Colorado at
  Boulder, \\ Campus Box 390, Boulder, CO 80309-0390, USA\\E-mail:
  uta.stoesslein@desy.de 
}


\twocolumn[\maketitle\abstract{Recent progress in the field
  of spin physics of high energy particle interactions is reviewed
  with particular emphasis on the spin structure functions as measured
  in polarized deep inelastic lepton-nucleon scattering (DIS).  New
  measurements are presented to obtain more direct information on the
  composition of the nucleon angular momentum, with results from
  semi-inclusive DIS accessing flavour-separated parton distribution
  functions (PDF) and with first data from hard exclusive reactions
  which may be interpreted in terms of recently developed
  generalizations of parton distribution functions (GPD).  Finally,
  experimental prospects are outlined which will lead to a further
  development of the virtues of QCD phenomenology of the spin
  structure of the nucleon.  
}]

\section{Introduction}
The understanding of strong interactions including spin as an
additional degree of freedom is an intensively discussed question
since Quantum Chromodynamics (QCD) became the gauge field theory of
the strong interaction establishing the intuitive quark model as the
valid concept for the nucleon substructure.

The quark-parton-model (QPM) could successfully explain the positive
cross section asymmetries observed in deep inelastic polarized
electron-proton experiments performed early at
SLAC~\cite{us:e80,us:e130}.  The modern age in QCD spin physics began
when at CERN a high energy muon beam experiment accessed spin
phenomena in an extended kinematic range of the four-momentum transfer
squared $Q^2$ and Bjorken-$x$.  The well-known result was the
discovery, made in 1987 by the EMC experiment~\cite{us:emc}, that only
a small fraction, $\frac{1}{2}\Delta \Sigma=\frac{1}{2}(0.12 \pm
0.17)$, of the proton's spin is due to the spin of quarks, contrary to
expectations of the naive QPM ($\Delta \Sigma=1$) or the relativistic
QPM ($\Delta
\Sigma=0.58$).  Perturbative QCD (pQCD) in next-to-leading order (NLO)
was able to attribute the small value of $\Delta \Sigma$ to the axial
anomaly in the polarized photon-gluon-scattering cross section, which
depends on the factorization and renormalization scheme.

Meanwhile a wealth of experimental data using deep inelastic
scattering of polarized lepton beams off polarized targets has been
taken at
SLAC~\cite{us:E142_96a,us:E143_98a,us:E154_97a,E155_99a,us:E155_00a},
CERN~\cite{us:SMC_98a} and DESY~\cite{us:Hermes_97a,us:Hermes_98a}
which confirms $\Delta \Sigma$ to be small with typical values of
about 0.2-0.4.  The old question thus remains unanswered yet: Where is
the remainder of the nucleon spin?  Is it measurable and calculable?
How important are non-perturbative effects of QCD which are
immediately related to low energy physics and confinement?  Which role
are contributions from gluons or sea quarks playing as well as the orbital
angular momentum of quarks and gluons?  Are there new observables which
could help to answer some of these questions?

In this report the recent progress in QCD spin physics is summarized,
highlighting new precise data for the spin structure functions $g_1$
and $g_2$ from SLAC, Hermes and JLAB, updated estimates of polarized
quark and gluon polarizations from NLO pQCD evolution, and more
directly from interpreting semi-inclusive Hermes data in LO pQCD.
First data from JLAB and Hermes on hard exclusive processes are
presented here, which potentially represent a unique experimental
input to be used in conjunction with further polarized DIS data to
identify the quark orbital angular momentum contribution to the
nucleon spin.

New single-spin azimuthal asymmetry measurements from Hermes are
presented. These suggest the yet unknown leading twist transversity
distribution functions are accessible using transversely polarized
targets at Hermes and protons at RHIC in the near future.  Spin
physics in polarized pp-collisions at RHIC opens new ways to determine
the gluon polarization and also flavour separated quark spin
distributions, complementary to data expected to come from the COMPASS
experiment at CERN and from Hermes at DESY.
\section{Experiments and Kinematics} \label{us:secexp}
\begin{table*}
\begin{footnotesize}
\begin{center}
\begin{tabular}{|c|c|c|c|c|c|c|c|}
\hline
 { Lab} &  { Experiment }&  { Year }&  {Beam  } &
 {$P_{\mathrm B}$ }&  {Target} & {$P_{\mathrm T}$ } &  {$f$} \\ 
\hline
SLAC & E80~\cite{us:e80}   &   75 & 10-16 GeV $e^-$     & 0.85  & H-butanol & 0.50  & 0.13  \\
     & E130~\cite{us:e130}  &   80 & 16-23 GeV $e^-$     & 0.81  & H-butanol & 0.58  & 0.15 \\
     & E142~\cite{us:E142_96a}  &   92 & 19-26 GeV $e^-$   & 0.39  & $^3$He    & 0.35  & 0.35 \\
     & E143~\cite{us:E143_98a}   &   93 & 10-29 GeV $e^-$      & 0.85  & NH$_3$    & 0.70 & 0.15  \\
     &       &      &                      &      & ND$_3$    & 0.25   &0.24  \\
     & E154~\cite{us:E154_97a}  &   95 & 48 GeV $e^-$       &  0.83  & $^3$He    & 0.38  & 0.55  \\
     & E155~\cite{us:E155_00a}   &   97 & 48 GeV $e^-$       &  0.81  & NH$_3$    & 0.80  & 0.15 \\
     &       &      &                    &          & LiD       & 0.22  & 0.36  \\
     & E155X~\cite{us:E155X_01a} &    99& 29/32 GeV $e^-$       &  0.83  & NH$_3$    & 0.70 
& 0.16  \\
     &       &      &                                &      & LiD       & 0.22  & 0.36  \\
\hline
CERN & EMC~\cite{us:emc}  &    85 &200 GeV $\mu^+$ & 0.79  & NH$_3$    & 0.78  & 0.16  \\
     & SMC~\cite{us:SMC_98a}  &    92 &100 GeV $\mu^+$     & 0.81  & D-butanol & 0.40  & 0.19  \\
     &      &    93 &190 GeV $\mu^+$     & 0.80  & H-butanol & 0.86  & 0.12  \\
     &      & 94/95 &                    &  0.80  & D-butanol & 0.50  & 0.20  \\
     &      &    96 &                     & 0.80  & NH$_3$    & 0.89  & 0.16 \\
\hline
DESY &HERMES~\cite{us:Hermes_97a,us:Hermes_98a,us:lenisa}
     &    95 & 28 GeV $e^+$       & 0.55  & $^3$He    & 0.46  & { {1.0}} \\
     &      & 96/97 &                    & 0.55  & H         & 0.88  &  { {1.0}} \\
     &      & 98    & 28 GeV $e^-$       & 0.55  & D         & 0.85  & { {1.0}} \\
     &      & 99/00 & 28 GeV $e^+$       & 0.55  & D         & 0.85  &  { {1.0}} \\
\hline
  { DESY }&   {HERMES~\protect\cite{us:vincter} }&   { $\geq$01} & 28 GeV $e^{\pm}$
& 0.55  & H    
    & 0.85  & 1.0 
\\
\hline
 { CERN} &  { COMPASS~\protect\cite{us:compassprop,us:kabuss}}&    { $\geq$01} &160 GeV $\mu^+$     & 0.80  & NH$_3$    & 0.90  & 0.16 \\
     &       &      &                    &      &  LiD      & 0.50  & 0.50  \\
\hline
  {BNL } &   RHIC~\cite{us:rhic} &     { $\geq$01} &  {200 GeV p }     &
0.70  &   {200 GeV p}  & 0.70    & { {1.0}} \\
\hline
\end{tabular}
\caption{High energy spin physics experiments.
} 
 \label{usfig:exp}  
\end{center}
\end{footnotesize}
\end{table*} 
The high energy facilities and experiments investigating the nucleon
spin structure using polarized charged lepton beams and polarized
targets are summarized in Tab.~\ref{usfig:exp},
where typical beam and target parameters are listed.  The values
achieved reflect the enormous development in target and beam
technology over the last three decades.  Besides the high values for
the beam and target polarizations, of about 0.5 to 0.9, the factor $f$
enters additionally as a dilution of polarization dependent
quantities; it denotes the fraction of polarizable target material.
For typical solid state target materials, as used at SLAC and CERN,
$f$ is small and varies from about 0.13 for butanol to 0.5 for lithium
deuteride and is about 0.55 for a $^3$He gas target (E154).  The
Hermes experiment has an internal target with almost pure gases where
$f$ is close to unity.  The polarization as well as the $f$ values, in
conjunction with beam intensity and target thickness determine the
statistical precision, which is largest for the E154/E155 experiments
at the high-intensity 48 GeV SLAC electron beam and lower for SMC.
The high beam energy of the CERN muon beam provided access to low
Bjorken-$x$ ($x<0.01$) and to larger values of $Q^2$
($Q^2>10$~GeV$^2$).

The kinematic coverage of the fixed target experiments is essentially
determined by the incoming beam energy $E$ defining the squared
centre-of-mass energy $s = 2ME$ which limits the maximum $Q^2 = 4 E E'
\sin^2 \theta/2$ to $Q^2_{\mathrm {max}} = s$.  The scaling variables
Bjorken-$x$ and inelasticity $y$ are given by $ x={Q^2}/{2M\nu}$ and
$y=\nu/E$, where $M$ is the nucleon mass and $\nu = E - E'$ is the
energy transferred by the virtual photon to the target nucleon.  The
energy $E'$ and the polar angle $\theta$ of the scattered charged beam
lepton determine the DIS kinematics.  To cover also the higher $Q^2$
values, the E155/E155X experiments were performed with three
independent magnetic spectrometers at central angles of 2.75$^\circ$,
5.5$^\circ$, and 10.5$^\circ$.  The SMC and Hermes spectrometers have
a large acceptance in the forward direction which allows for
semi-inclusive measurements.

The experiments complement each other in their sensitivity to possible
systematic uncertainties associated in particular with the beam and
target polarization; these are typically controlled at the level of 2
to 5\%.  For instance, different attempts were made to minimize
instrumental asymmetries: SMC used two target cells with opposite
polarization simultaneously, while the polarization of the Hermes gas
target, which can be inverted within milliseconds, was flipped in
cycles. At SLAC the helicity for each beam pulse was randomly
selected.  Whereas SLAC experiments and SMC determined the
polarization of the target material $P_{\mathrm T}$ by nuclear
magnetic resonance measurements, Hermes is using a
Breit-Rabi-Polarimeter (for p,d).  Also the beam polarimetry is rather
different. The beam polarization $P_{\mathrm B}$ was determined using
M{\o}ller scattering from polarized atomic electrons at SMC and SLAC
experiments, but at Hermes utilizing the spin dependence of Compton
backscattering of circularly polarized photons off polarized
electrons.

In 2001, with COMPASS at the CERN high-intensity muon beam line M2 and
Hermes Run II at DESY, two powerful fixed target experiments are 
operational.  The recent commissioning of the polarized proton rings at
RHIC is an important milestone in experimental spin physics.  Two
large collider detectors, PHENIX~\cite{phenix} and STAR~\cite{star},
along with several smaller experiments, will participate in the RHIC
spin programme.

\section {Asymmetries and Structure Functions $g_{1}$ and $g_{2}$ }
Measurements of polarized nucleon structure functions in inclusive DIS
have been traditionally a prime focus of spin physics experiments.
Using longitudinally polarized charged lepton beams and polarized
targets the helicity-dependent quark distributions can be probed. This
sensitivity results from angular momentum conservation, i.e.  virtual
photons can only be absorbed by a quark if their spins are oriented
antiparallel.  In the naive QPM helicity-averaged and
helicity-dependent quark momentum distribution functions are
introduced
\bea
q(x) & = & q^{+}(x)+q^{-}(x), \\
\Delta q (x)  & = & q^{+}(x)-q^{-}(x),
\eea
which are directly related to the spin-averaged structure 
function $F_1$ and the spin-dependent structure 
function $g_1$ 
\bea
F_1(x) & = & \frac{1}{2}\sum_{q} e_q^2 q(x), \\
g_1(x) & = & \frac{1}{2}\sum_{q} e_q^2 \Delta q(x). 
\eea
Here the sums extend over both quark and anti-quark flavors weighted
by the electrical charge $e_q$ squared.  The notations $q^{+(-)}$
refer to parallel (antiparallel) orientation of the quark and nucleon
spins.  The second leading twist spin structure function $g_2$
measurable in DIS is expected to be zero in the naive QPM.

The experimentally asymmetries accessible with a longitudinally
polarized charged lepton beam are defined as~\footnote{Here the factor
$1/P_{\mathrm B} P_{\mathrm T} f$ which accounts for dilution and for
beam and target polarizations, see Sec.~\protect\ref{us:secexp}, is
set to unity for the sake of simplicity.}
\bea 
A_{||} =  
 \displaystyle\frac{\sigma^{\stackrel{\rightarrow}{\Leftarrow}}-\sigma^{\stackrel{\rightarrow}{\Rightarrow}}}
{\sigma^{\stackrel{\rightarrow}{\Leftarrow}}+\sigma^{\stackrel{\rightarrow}{\Rightarrow}}},
\:
A_{\perp}  =    \displaystyle\frac{\sigma^{\scriptstyle\Downarrow  \scriptstyle\rightarrow} - \sigma^{\scriptstyle\Uparrow  \scriptstyle\rightarrow}}{
\sigma^{\scriptstyle\Downarrow  \scriptstyle\rightarrow} + \sigma^{\scriptstyle\Uparrow  \scriptstyle\rightarrow}} ,
\eea
where $\sigma^{\stackrel{\rightarrow}{\Rightarrow}}$
($\sigma^{\stackrel{\rightarrow}{\Leftarrow}}$) is the cross section
for the lepton and nucleon spins aligned parallel (anti-parallel),
while $\sigma^{\scriptstyle\Downarrow \scriptstyle\rightarrow}$
($\sigma^{\scriptstyle\Uparrow \scriptstyle\rightarrow}$) is the cross
section for nucleons transversely polarized w.r.t. the beam spin
orientation.  These cross section asymmetries are related to two
virtual photon asymmetries, $A_1$ and $A_2$, through %
\bea 
A_{||} =  D\left( A_1 + \eta A_2\right), \,
 A_\perp =  d\left(A_2 - \zeta A_1\right).  
\eea 
The factors $D$ and $d$ denote the virtual
photon polarization where $D\approx y$.  They are explicitly given by
$D = [1-(1-y)\epsilon]/(1+\epsilon R)$, which depends on the ratio of
longitudinal to transverse virtual-photon absorption cross sections
$R=\sigma_{\rm L}/\sigma_{\rm T}$, and by $d = D \sqrt{{2\epsilon /(
    1+\epsilon)}}$.  The quantity $\epsilon = [4(1-y) -\gamma^2
y^2]/[2y^2+4(1-y)+\gamma^2 y^2]$ describes the flux ratio of
longitudinal to transverse photons.  The kinematic factors are defined
as $\eta = \epsilon \gamma y/[1-\epsilon(1-y)]$,
$\gamma=2Mx/\sqrt{Q^2}$ with $ \eta \approx \gamma \:\:
^{\underrightarrow{Q^2 \gg M^2} } \:\: 0$, and $\zeta =
\eta\left({1+\epsilon / 2\epsilon}\right)$.

The virtual photon asymmetries are bound by
$|A_2|\le\sqrt{R(1+A_1)/2}$~\cite{us:soffer00} and $|A_1|\le 1$.  They
can be described in terms of the
spin-dependent structure functions %
\bea  
\label{useq:g1g2}
A_1 & = & \displaystyle\frac{g_1-\gamma^2 g_2}{F_1}\approx \frac{g_1}{F_1}, \\
 A_2& = & \gamma \displaystyle\frac{g_1+g_2}{F_1},          
\eea 
where $F_1=F_2(1+\gamma^2)/2x(1+R)$ can be determined from
measurements of $R$ and of the well-known unpolarized structure
function $F_2$.  From Eq.~\ref{useq:g1g2} one can deduce that use of
longitudinally polarized target predominantly determines $g_1$, while
DIS experiments using a transversely polarized target are sensitive to
$g_1+g_2$.

\begin{figure*}
  \includegraphics[width=\figsize]{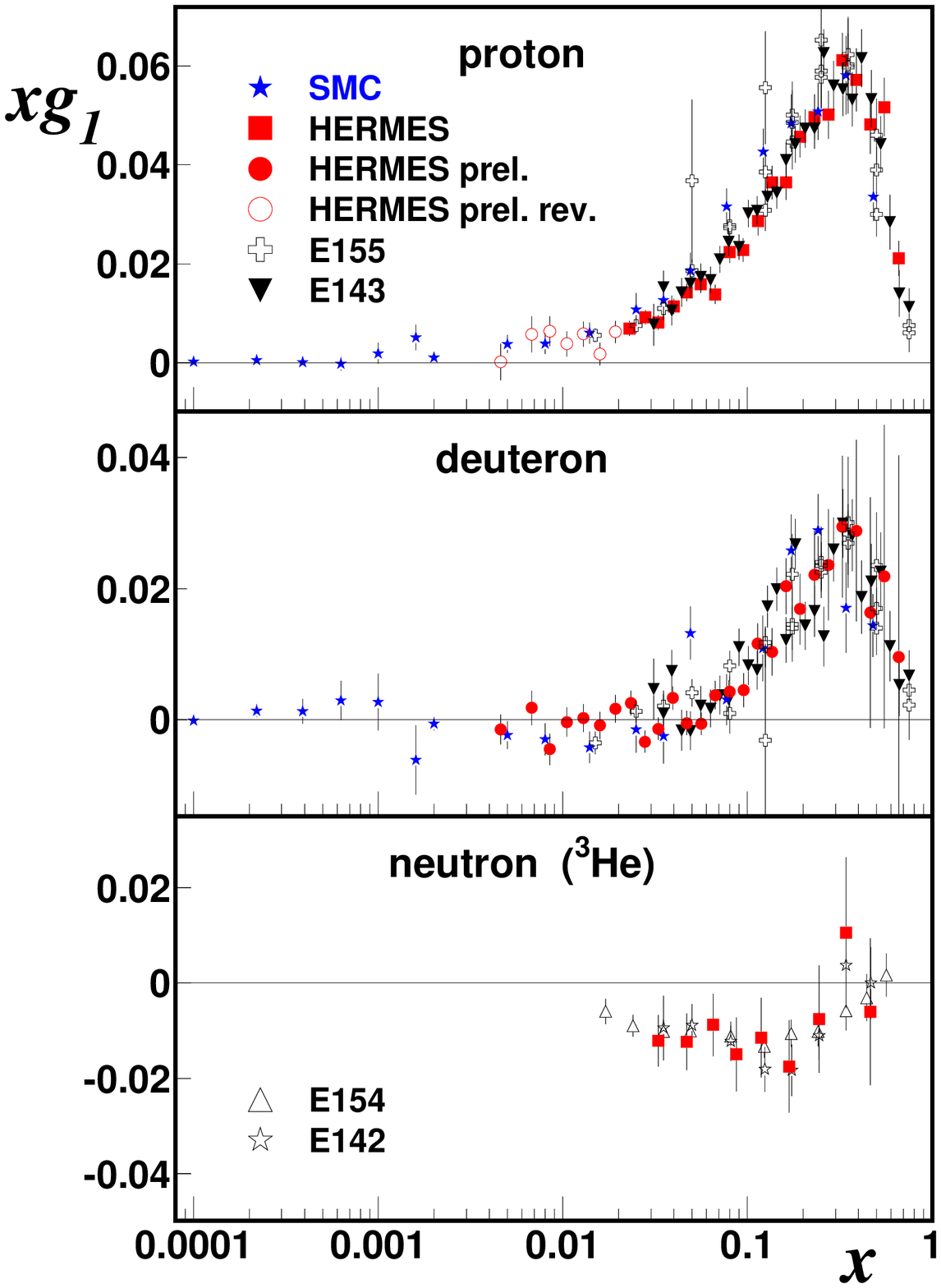} \hspace*{2mm}
  \includegraphics[width=\figsize]{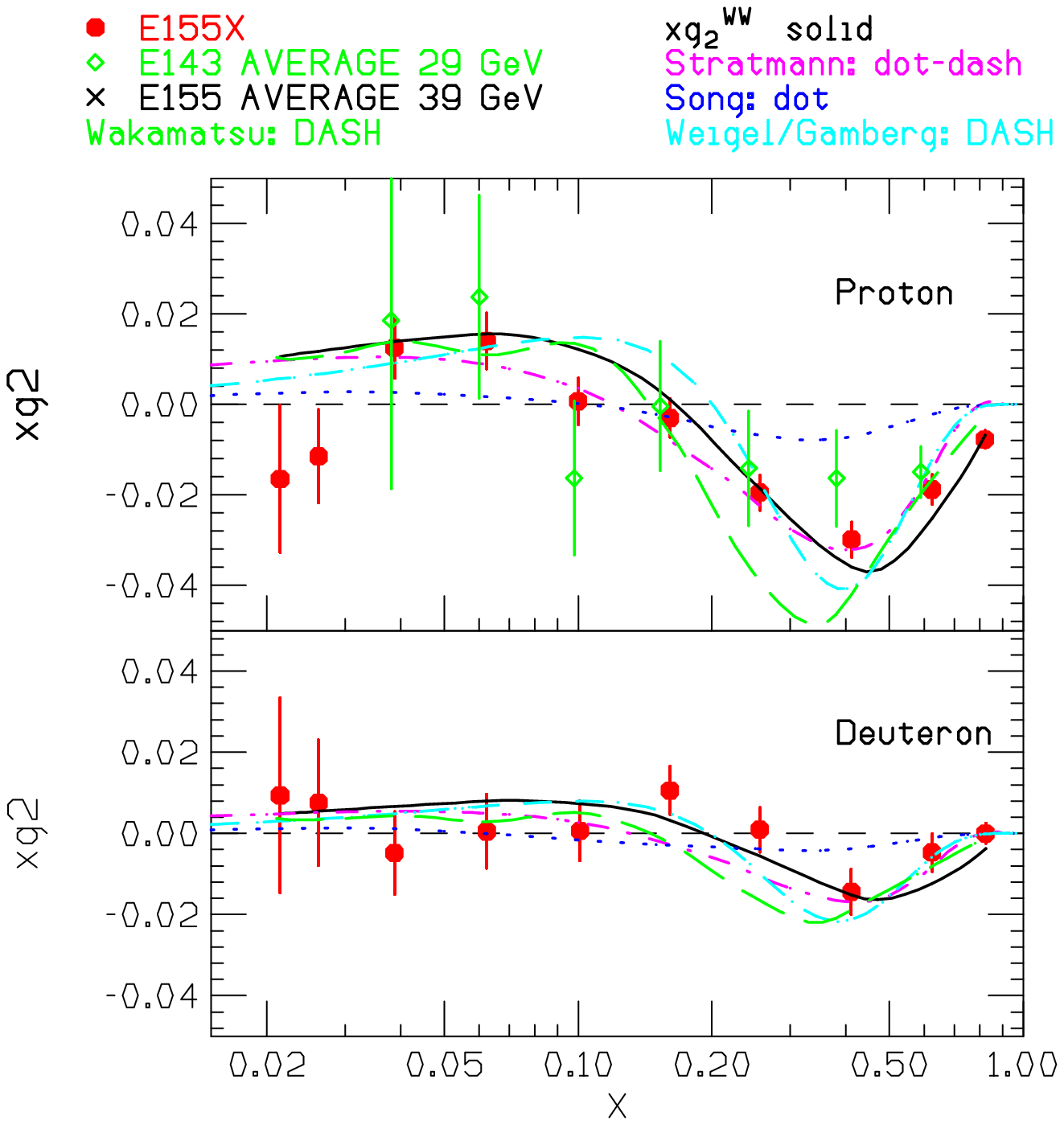}
\caption{Compilation of recent  data on the spin structure functions 
  $xg_1(x, Q^2) $ (left) and $xg_2(x, Q^2)
  $~\protect\cite{us:E155X_01a} (right) including new data from HERA
  (Hermes preliminary) and SLAC (E155X). All data are given at their
  quoted mean $ Q^2$ values.  The values for $xg_2 $ are compared with
  the $g_2^{\mathrm WW}$ term (solid line), see text,  and several bag
  model calculations~\protect\cite{us:E155X_01a}.
  \label{usfig:g1world}} 
\end{figure*} 
The extraction of the proton structure function $g_1^{\mathrm p}$ from
a longitudinally polarized hydrogen target is straightforward.
Extraction of the neutron structure function $g_1^{\mathrm n}$ from
longitudinally polarized $^2$H or $^3$He targets, however, requires
additional nuclear corrections to be applied.  The polarization of the
$^3$He nucleus is mainly due to the neutron.  Corrections due to the
nuclear wave function of the polarized $^3$He nucleus have to be
applied using $g_1^{\mathrm p}$ data to evaluate the proton
contributions, i.e.,
\begin{equation} 
g_1^{\mathrm n}(x,Q^2) = \displaystyle\frac{1}{\rho_{\mathrm n}}(g_1^{^3{\mathrm {He}}} - 2 \rho_{\mathrm p} g_1^{\mathrm p}) \, , 
\end{equation}
where $\rho_{\mathrm n} = (0.86\pm 0.02)$ and $\rho_{\mathrm
  p}=(-0.028\pm 0.004)$ are taken from a number of
calculations~\cite{us:friar,us:nuclcorr1}. Additional corrections due
to the neutron binding energy and Fermi motion were shown to be
small~\cite{us:nuclcorr2}.
For the polarized deuteron, the large contribution due to the
polarized proton must be subtracted. In addition the D-state component
in the deuteron wave function will slightly reduce the deuteron spin
structure function due to the opposite alignment of the p-n spin
system in this orbital state.  This leads to 
\begin{equation} 
g_1^{\mathrm n}(x,Q^2) =
\displaystyle\frac{2 g_1^{\mathrm d}(x,Q^2)}{ (1 - 1.5 \omega_{\mathrm D})} -
g_1^{\mathrm p}(x,Q^2)\, , 
\end{equation}
where $\omega_{\mathrm D} = 0.05\pm 0.01$~\cite{us:lac81} is the
D-state probability of the deuteron.  Furthermore, a possible
contribution may occur from the still unknown tensor-polarized
structure function $b_1^{\rm d}$, a new leading twist function that
occurs in case of scattering electrons off a spin-1
target~\cite{us:hood89}. In the QPM, $b_1$ measures the difference in
the quark momentum distributions of a helicity $1$ and $0$ target,
i.e.  $b_1 = \frac{1}{2}(2q^0_\uparrow - q^1_\uparrow -
q^1_\downarrow)$. Here $q^m_\uparrow$ ($q^m_\downarrow$) is the
probability to find a quark with momentum fraction $x$ and spin up
(down) in a hadron or nucleus with helicity $m$.  In all analysis so
far $b_1^{\rm d}=0$ is assumed, but first data to constrain $b_1^{\rm
  d}$ are expected from Hermes~\cite{us:us00}.

The recent world data on the spin structure functions $xg_1(x, Q^2) $
and $xg_2(x, Q^2) $ for the proton, deuteron and neutron are presented
at their measured $\langle Q^2 \rangle$ values in
Fig.~\ref{usfig:g1world}.  Also shown are the final high-precision
data from E155, and new preliminary data from Hermes in the kinematic
range $0.002 <x<0.85$ and 0.1~GeV$^2<Q^2<20$~GeV$^2$, where the data
at small Bjorken-$x<0.01$ belong to low photon virtualities of
$Q^2<1.2$~GeV$^2$.  These Hermes data confirm the { SMC} small-$x$ and
$Q^2>1$~GeV$^2$ data for $A_1^{\rm p,d} $ for the first time.  Even
lower $x$ values down to $x=6\cdot10^{-5}$ but at extremely low
$Q^2=0.01$~GeV$^2$ were reached by SMC with a dedicated low-$x$
trigger~\cite{us:SMC_99a}.

According to Fig.~\ref{usfig:g1world} the spin structure functions
$g_1 $ and $g_2 $ are best known for the proton. The precision and
kinematic coverage of the $g_1 $ data is much better than for the $g_2
$ data.  This holds even with the factor of three improved
measurements of the dedicated E155X runs as compared to previous $g_2
$ data from SMC~\cite{us:g2smc} or from SLAC collaborations
E142~\cite{us:E142_96a}, E143~\cite{us:g2e143}, E154~\cite{us:g2e154}
and E155~\cite{us:g2e155}.

 A comparison of  $xg_1 $ and  $xg_2 $ values
leads to the striking observation that at high $x\sim 0.4 $ 
the values of  $g_1$  
are positive and of $g_2$ are negative, but different from zero for
both the proton and the deuteron.  For $g_2$, with so far limited
accuracy, this is in contrasts to the parton model expectation of
$g_2=0$. There exists no simple partonic picture for $g_2\ne 0$ which
can be understood in terms of incoherent scattering of massless,
collinear partons.  It turns out that $g_2$ arises from
higher-twist processes which can be described in terms of coherent
parton scattering.  The spin structure function $g_2$ is thus an
interesting example for a higher-twist observable representing
quark-gluon correlations in the nucleon.
\begin{figure} 
\begin{center}
\includegraphics[width=\figsizec]{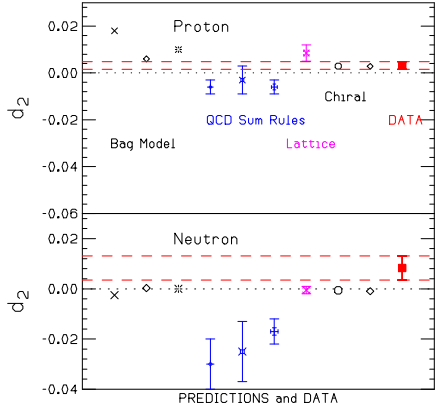}
\caption{Calculations of the twist-3 matrix element $d_2$ of 
  the proton and the neutron.  The dashed lines indicate the present
  uncertainty in the data.  Also shown are bag model, chiral quark
  model, QCD sum rule and recent QCD lattice
  calculations~\protect\cite{us:E155X_01a}.
  \label{usfig:d2}} 
\end{center}   
\end{figure}

In fact, omitting quark mass terms, $g_2$ can be described by a twist-two
contribution, the so-called $g_2^{\rm WW}$
Wandzura-Wilczek-term~\cite{us:ww} calculable from $g_1$, and a pure
twist-3 term denoted here with $\bar {g_2}$ reflecting the
interaction-dependent part,
\begin{eqnarray} 
{g_2}(x,Q^2)&=&{g_2^{\rm WW}}+{\bar g_2}(x,Q^2),\\
{g_2^{\rm WW}}(x,Q^2)&=&-g_1(x,Q^2)+ \nonumber \\
 & & + \int_x^1\frac{g_1(y,Q^2)}{y}\,{\rm d} y. 
\label{useq:ww}
\end{eqnarray} 
The recent data from the $g_2$ experiment E155X (right panel of
Fig.~\ref{usfig:g1world}) are in good agreement with the
Wandzura-Wilczek approximation, $g_2^{WW} \propto -g_1$, which
supports the observation discussed above.  Consequently, any possible
twist-3 term must be small. This is supported by new
estimates~\cite{us:E155X_01a} of the twist-3 matrix element $d_2$,
\begin{equation}
d_2(Q^2) = 3 \int_0^1 x^2 \, [ g_2(x,Q^2) - g_2^{\rm WW}(x,Q^2) ]\,{\rm d} x \, ,
\end{equation}
obtained at an average $Q^{2} $ of 3 GeV$^{2}$ with an improved
precision for the proton and the neutron,
$d_2^{\mathrm p}=0.0032\pm0.0016$ and $d_2^{\mathrm
  n}=0.0083\pm0.0048$, respectively.  Here, the $ g_2^{\rm WW}$-term
was calculated using empirical fits~\cite{us:E155_00a} to $g_1$ data.
The present status of $d_2$ measurements and model calculations is
shown in Fig.~\ref{usfig:d2}.  Further details about the complex
nature of $g_2$ and $d_2$ may be found e.g. in
Ref.~\cite{us:ji_overview}.

\section {QCD Analyses and Gluon Polarization $\Delta G$}

QCD predicts the $Q^2$ dependence of unpolarized and polarized
structure functions measured in DIS.  Before discussing the
theoretical framework and pQCD analyses of $g_1$, it is worthwhile to
look at the experimental status of measuring the $Q^2$ dependence of
spin structure functions.

\begin{figure*} 
\begin{center}
\includegraphics[width=\figsize]{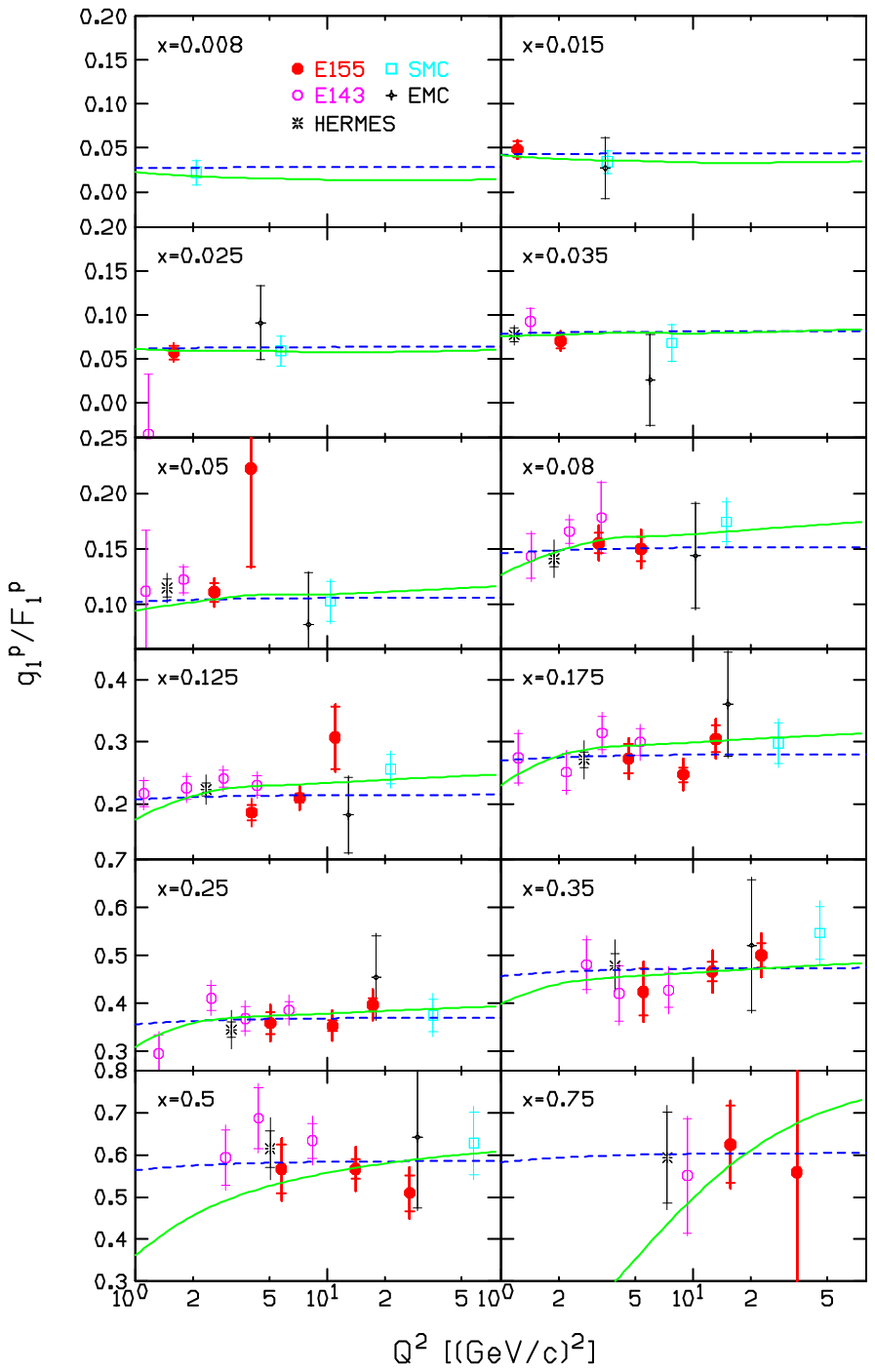}
\hspace*{3mm}
\includegraphics[width=\figsize]{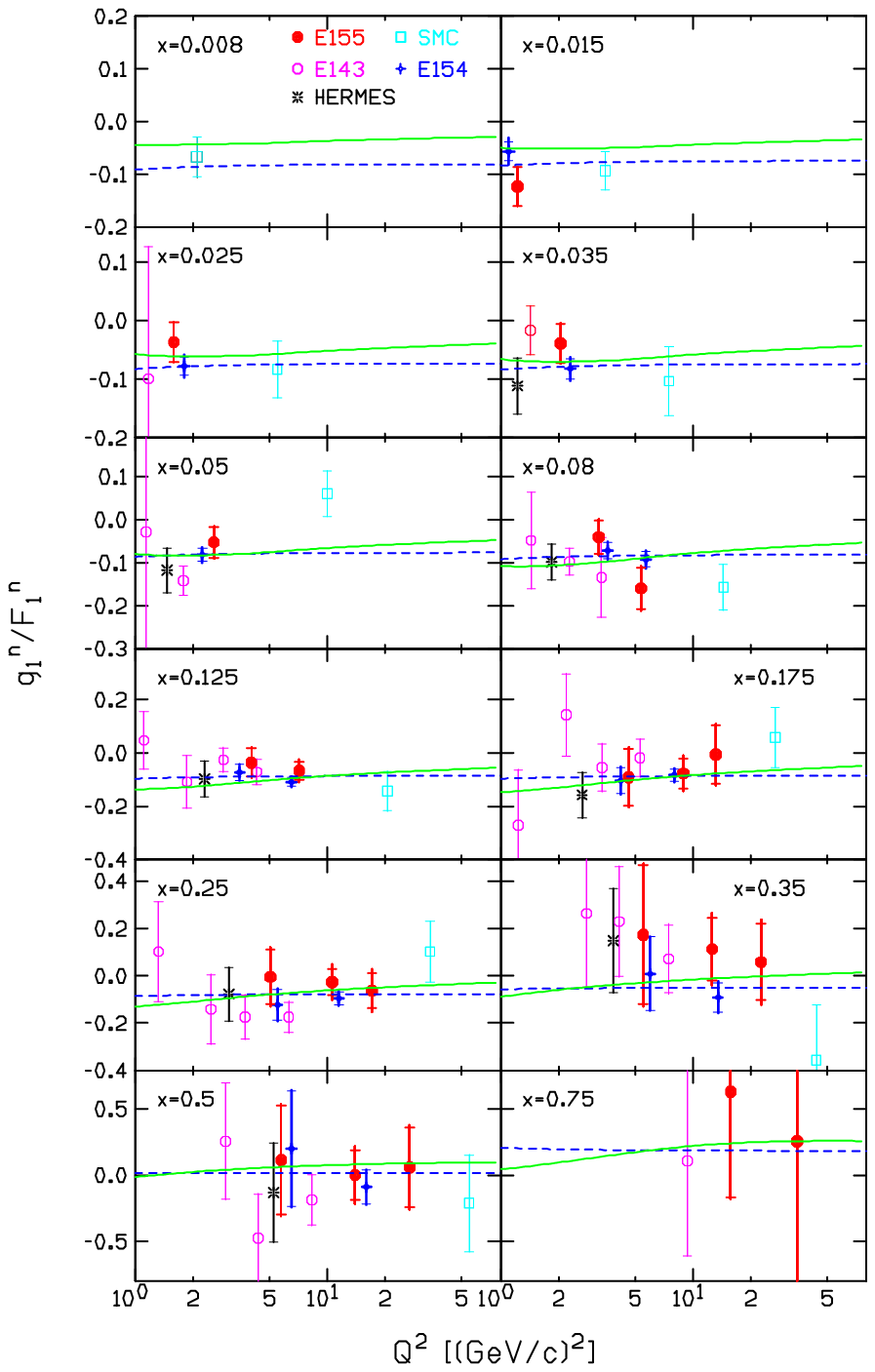}
\caption{The $Q^2$ dependence of $g_1/F_1$, 
  note $ g_1/F_1 \approx A_1$, of the proton (left) and neutron (right)
  for $Q^2>1$~GeV$^2$ data. Also shown are the phenomenological fits
  (dashed line) and results from the E155 pQCD fit in NLO (dotted 
  line)~\protect\cite{us:E155_00a}.
  \label{usfig:g1f1q2}} 
\end{center}
\end{figure*}

A compilation~\protect\cite{us:E155_00a} of recent DIS data of the
structure function ratio $g_1/F_1$ is shown in
Fig.~\ref{usfig:g1f1q2}.
In any given $x$ bin, there is no experimental evidence of a strong
$Q^2$ dependence of $g_1/F_1$. A phenomenological fit to recent data
with $Q^2>1$ GeV$^2$ and an energy of the hadronic final state $W> 2$
GeV, parametrizes a possible $Q^2$ dependence
by~\protect\cite{us:E155_00a}
\begin{eqnarray}
\displaystyle\frac{g_1^{\mathrm p}} {F_1^{\mathrm p}}&=&x^{0.700}(0.817+1.014x-1.489x^2) \nonumber \\
 & & \times (1+\displaystyle\frac{c^{\mathrm p}}{Q^2}) \label{useq:g1f1e155}\\
\displaystyle\frac{g_1^{\mathrm n}} { F_1^{\mathrm n}}&=&x^{-0.335}(-0.013 -0.330x+0.761x^2)\nonumber \\ 
& & \times (1+\displaystyle\frac{c^{\mathrm n}}{Q^2}).
\end{eqnarray}
The coefficients $c^{\mathrm p} = -0.04\pm0.06$ and $c^{\mathrm
  n}=0.13\pm0.45$ describing the $Q^2$ dependence found to be small
and consistent with zero.

The transition from DIS to the low $Q^2$ regime, $Q^2 \lesssim
1$~GeV$^2$, and to the resonance region $W \lesssim 2$~GeV is
investigated at TJNAF with beam energies of 0.8-5.7~GeV.  Using
polarized electrons ($P_{\mathrm B}\sim 70$\%) scattered off polarized
solid state targets significant, complementary measurements of $g_1$
and $g_2$ for $0.02<Q^2<1.2$~ GeV$^2$ are anticipated: first
preliminary results for the asymmetry $A_1 + \eta A_2$ are of
remarkable statistical precision. The asymmetries show strong $Q^2$
dependences for a $^3$He target~\cite{us:roblin} as well as for proton
(NH$_3$) and deuteron (ND$_3$) targets~\cite{us:dodge}.  This behavior
reflects the rapidly changing helicity structure of some resonances
with the scale probed.  Accurate measurements will allow stringent
tests of nucleon structure models and may shed new light to the
important question at which distance scale pQCD corrections will break
down and physics of confinement may dominate.

\begin{figure} 
\begin{center}
\includegraphics[width=\figsize]{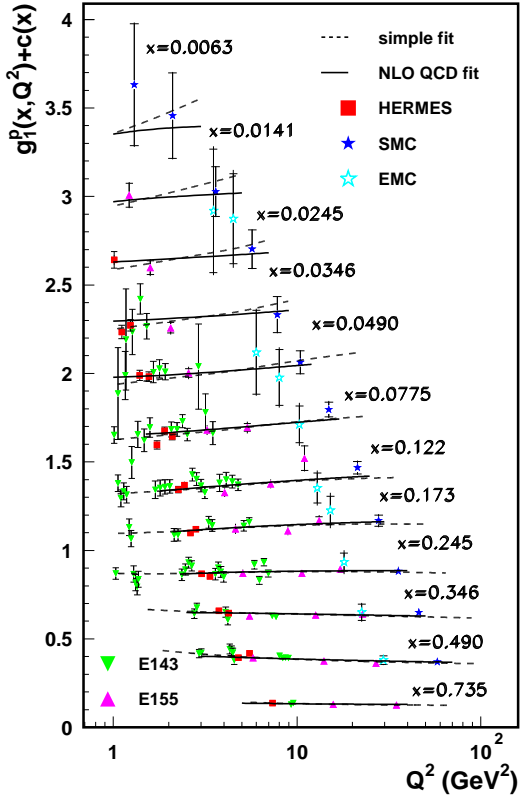}
 \caption{$Q^2$ dependence of $g_1^\mathrm{p}(x, Q^2)$ 
   for $Q^2>1$~GeV$^2$. The $g_1^{\rm p}$ values are calculated using
   recent data on ratio $g_1^{\rm p}/F_1^{\rm p}$ and the function
   $F_1^{\rm p}$ determined by $F_2^{\rm p}$~\protect\cite{us:allm97}
   and $R$~\protect\cite{us:withlow}.  To evaluate the $Q^2$ behavior,
   the data has been shifted to common $x$ values using $F_1^{\rm p}$
   and $g_1^{\rm p}/F_1^{\rm p}$ parameterizations.  Also shown are
   the fit according to Eq.~\protect\ref{useq:g1f1e155} and a NLO pQCD
   fit~\protect\cite{us:BB}.
  \label{usfig:lnq2}}  
\end{center}
\end{figure}

While those tasks are still theoretically challenging, pQCD delivers a
good description of the $g_1$ data with $Q^2>1$~GeV$^2$.  This is
illustrated for the proton case in Fig.~\ref{usfig:lnq2}.  The
observed pattern of scaling violations of $g_1^{\rm p}$ resembles the
known pattern of scaling violation in the unpolarized structure
function $F_2^{\rm p}$.  The $Q^2$ evolution equations, given in the
DGLAP formalism, allow the unpolarized gluon distribution and the
strong coupling constant $\alpha_s$ to be determined~\cite{us:rev}.
Since gluons are vector particles they are expected to contribute to
the spin of the nucleon as well, asymptotically about 50\% of the
protons spin.

In NLO pQCD, the spin structure function $g_1$ is given by
\begin{eqnarray} \label{us:g1}
g_1^{\mathrm{p(n)}}  = \frac{1}{9} {\Bigg \{ }  \Delta  C_{NS}\otimes \left [
+ (-) \frac{3}{4} { \Delta q_3} + 
 \frac{1}{4} {\Delta q_8}  \right ]  \nonumber \\
 + \,{\Delta  C_S} \otimes { \Delta\Sigma} + 2 N_f { \Delta  C_G}
\otimes {\Delta G}   {\Bigg \} } ,\hspace*{4mm}
\end{eqnarray}
where $\Delta C_{q,G}$ are the spin-dependent Wilson coefficients and 
 $\otimes$ denotes the convolution in $x$ space. 
The usual notations  for three flavours ($N_f=3$) are: 
\begin{eqnarray} 
{\Delta\Sigma}&= &(\Delta u+\Delta \bar{u}) + (\Delta d+\Delta \bar{d}) + (\Delta s+\Delta \bar{s}) \nonumber \\
{ \Delta q_3} &= &(\Delta u+\Delta \bar{u}) - (\Delta
d+\Delta \bar{d}) = 6(g_1^{\mathrm p} - g_1^{\mathrm n}) \nonumber \\
{\Delta q_8} &= &(\Delta u+\Delta \bar{u}) + (\Delta
d+\Delta \bar{d}) -2 (\Delta s+\Delta \bar{s}) \nonumber \\
{ \Delta G}  &  & { \mathrm {in \, NLO  \, only}  } ;{ \Delta  C_G^0=0}
\mathrm { \:in \:LO}.  \nonumber
\end{eqnarray}
Apparently, inclusive data allow linear combinations of polarized
PDFs $(\Delta q + \Delta {\bar q})$ and the gluon polarization
$\Delta G$, as an ${\cal O} (\alpha_s) $ correction, to be accessed by
solving the DGLAP evolution equations.  The mixing of the evolution of
the quark singlet contribution $\Delta \Sigma $ and $\Delta G$ yields to 
 renormalization and factorization scheme dependent results in NLO
pQCD.  Frequently used schemes are the $\overline{\rm MS}$,
Adler-Bardeen (AB) or JET schemes, see e.g. Ref.~\cite{us:leader}.  In
the AB and JET schemes, $\Delta \Sigma $ is conserved and defined to
be a scale-independent quantity, which is related to the
$\overline{\rm MS}$ result by 
\bea 
\Delta \Sigma (Q^2)_{\overline{\mathrm{MS}}}  & = &   \Delta
  \Sigma_{\mathrm {AB(JET)}} \nonumber \\ 
& &  - N_f \frac{\alpha_s(Q^2)}{2 \pi} \Delta G(Q^2).  
\eea
The polarized gluon distribution is the same in all these schemes, $
\Delta G(Q^2)_{\overline{\mathrm{MS}}} = \Delta
G(Q^2)_{\mathrm{AB(JET)}}$. Observed differences of $\Delta G$ values
obtained might still point to systematic differences in the applied
theoretical descriptions, which includes e.g.  the treatment of quark
masses and flavours, the use of the strong coupling constant
$\alpha_s$ and the use of non-DIS data for further constraints,
respectively.

\begin{figure*} 
\begin{center}
\includegraphics[width=0.65\textwidth]{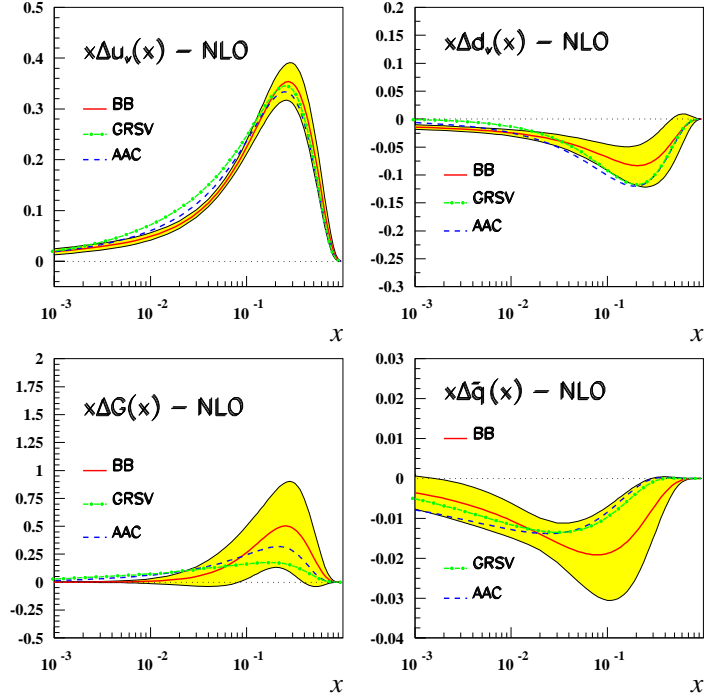}
 \caption{ Polarized parton distribution functions $x\Delta d_{\rm V}$, 
$x\Delta u_{\rm V}$, $x\Delta G$ and $x\Delta \bar{q}$ from updated
NLO pQCD ($\overline{\mathrm{MS}}$) fits 
at $Q^2 = 4 $~GeV$^2$ using SU(3)$_\mathrm{f}$  assumptions.  
Labels are according to the results from Ref.~\protect\cite{us:BB} (BB),
Ref.~\protect\cite{us:GRSV} (GRSV), and Ref.~\protect\cite{us:AAC} (AAC).
The shaded bands~\protect\cite{us:BB} 
 represent the propagated statistical errors only.
   \label{usfig:g1qcd1} }
\end{center}
\end{figure*}

Several groups have been performing spin-dependent NLO pQCD fits. The
one performed by the SMC collaboration was the first to carefully
treat statistical, systematic and theoretical
uncertainties~\cite{us:SMC_qcd}.  A new attempt to propagate the
statistical errors through the evolution procedure was done in
Ref.~\protect\cite{us:BB} and is presented in Fig.~\ref{usfig:g1qcd1}
together with other recent NLO fit results~\cite{us:GRSV,us:AAC}.  The
precise inclusive proton data predominantly constrain very well the
up-valence quark polarization to be positive and confirm together with
neutron and deuteron data the down-valence quark polarization to be
negative. The polarized sea is determined to be negative while $\Delta
G$ is suggested to be positive. However, both values have large
uncertainties as illustrated with the bands in
Fig.~\ref{usfig:g1qcd1}.

Spin-dependent pQCD analyses begin to become sensitive to the value of
$\alpha_s$ as well.  A recent determination of $\alpha_s$ yields
$\alpha_s({\mathrm{M}_\mathrm{z}^2}) = 0.114 \pm
0.005(\mathrm{stat})\,^{+0.009}_{-0.006}(\mathrm{scales})$~\cite{us:BB}
which is consistent with the value obtained by SMC,
$\alpha_s({\mathrm{M}_\mathrm{z}^2}) = 0.121 \pm
0.002(\mathrm{stat})\,{\pm 0.006}(\mathrm{syst\, and\,
thy})$~\cite{us:SMC_qcd}.

Present pQCD analyses use information from neutron and hyperon
$\beta$-decays to constrain the first moments  of the non-singlet distributions
($\Delta q_3$, $\Delta q_8$),
\begin{eqnarray}
a_3 & =&  \Delta q_3   =   F + D =  1.267 \pm 0.0035  \\
a_8 &  = &  \Delta q_8  =   3F - D =  0.585 \pm 0.025.  
\end{eqnarray}
Assuming SU(3)$_f$ flavour symmetry the first moment of $g_1$ is given
by
\bea
\Gamma_1(Q^2)   =  C_S(Q^2)a_0(Q^2) +  \hspace*{24mm}\nonumber \\
 + C_{NS}(Q^2)\frac{1}{12}\left(\left|\frac{g_a}{g_v}\right|-\frac{1}{3}(3F-D)\right),\hspace*{4mm}
\label{useq:ejsr}
\eea
where $\left|\ {g_a}/{g_v}\right|$ is the axial coupling constant and
$a_0(Q^2)$ is the axial charge.  An update of the
E154~\cite{us:E154fit} NLO pQCD fit in the $\overline{MS}$ scheme was
performed by the E155 collaboration\cite{us:E155_00a} using published
data. It further confirms the quark singlet contribution $\Delta
\Sigma$ to be small, $\Delta \Sigma=0.23\pm 0.04$(stat)$\pm 0.06
$(syst) at $Q^2=5$ GeV$^2$, well below the Ellis-Jaffe prediction
\cite{us:ellisjaffe} of 0.58.  The value for $\Gamma_1^{\mathrm p} - \Gamma_1^{\mathrm n }
=0.176\pm0.003\pm 0.007$ is found to be in agreement with the Bjorken
sum rule prediction of $0.182\pm0.005$.  For the first moment of the
gluon distribution a value of $\Delta G=1.6 \pm 0.8$(stat)$ \pm
1.1$(syst) is obtained, reflecting the fact, that the uncertainty on
$\Delta G$ from scaling violations is still too large to significantly
constrain the gluon contribution to the nucleon spin.  The value of
$a_8$ depends on the assumption of SU(3)$_f$ flavour symmetry among
hyperons, which is known to be inexact.  Indications of a breakdown of
SU(3) flavor symmetry, as observed in the hyperon $\beta$ decay, and
its impact on the polarized PDFs were discussed recently using the JET
scheme~\cite{us:leader01}. While the influence on the singlet and
non-strange quark polarizations was found to be small, the strange sea
quark and gluon polarizations change significantly when SU(3)$_f$
symmetry breaking effects are considered~\cite{us:leader01}: e.g.
$\Delta s+\Delta \bar{s}$ varies from $-0.02$ to $-0.15$ and $\Delta 
G$ from 0.13 to 0.84.  The observed strong dependence of gluon and
strange quark polarizations on the SU(3)$_f$ symmetry assumptions
calls for more direct probes for a determination of those quantities.
\begin{figure} 
\begin{center}
\includegraphics[width=\figsizeb]{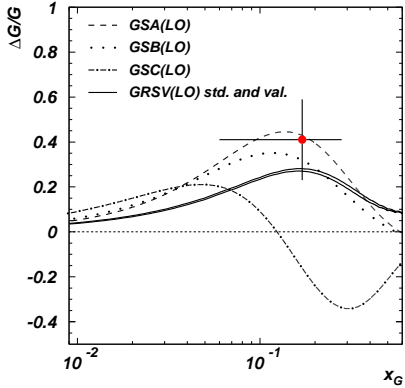}
 \caption{ Hermes result~\protect\cite{us:Hermesgluon} for $\Delta
 G/G$ extracted from high $p_T$ unlike sign hadron pair production.
 The result is compared to LO pQCD fits to a subset of the world's
 data on $g_1$: curves are from
 Refs.~\protect\cite{us:GSgluon,us:grsvgluon} evaluated at $Q^2 =
 2$~GeV$^2$.  The error bar on $\Delta G/G$ represents statistical and
 experimental systematic uncertainties only; no theoretical
 uncertainty is included.  \label{usfig:hermesgluon} }
\end{center}
\end{figure} 

A first attempt to determine $\Delta$G$/$G from the photon-gluon
fusion process (PGF) was presented by
Hermes~\protect\cite{us:Hermesgluon}.  Using a polarized hydrogen
target and selecting events with two hadrons with opposite charge and
high transverse momentum $p_T$, the double-spin asymmetry and its
dependence on $p_T^{h_1}$ and $p_T^{h_2}$ was measured.  For $h_1h_2$
pairs with p$_T^{h_1}>1.5$ GeV and p$_T^{h_2}>1.0$ GeV the asymmetry
is $A_{||}=-0.28\pm 0.12~{\rm (stat)}\pm 0.02~{\rm (syst)}$.  The
result has been interpreted considering contributions from deep
inelastic scattering, vector meson dominance and the two direct
leading order QCD processes: PGF and QCD Compton scattering.  Using
the PYTHIA~\cite{us:pythia} program the relative cross section
contributions were determined and the negative asymmetry explained by
a positive gluon polarization.  The asymmetry of the QCD Compton
process was calculated to be positive whereas the asymmetries of all
other subprocesses were assumed to be zero.  The extracted value of
$\langle\Delta G/G\rangle=0.41 \pm 0.18$$~{\rm (stat)}$$\pm$$
0.03~{\rm (syst)}$ is compared in Fig.~\ref{usfig:hermesgluon} to
several phenomenological LO pQCD fits where the hard scale of this
process is given rather by $\langle {p}_T^2 \rangle=2.1$~GeV$^2$ and
not by $\langle Q^2\rangle$ of $0.06$~GeV$^2$.  The positive sign of
$\langle\Delta G/G\rangle$ at $\langle x_G\rangle =0.17$ can only be
altered by a large negative spin asymmetry from some neglected
process, other than PGF.

A similar $\Delta G/G$ analysis is being performed by Hermes based on
the high statistics deuteron data of the run periods
1998-2000~\cite{us:aschenauer}.  The interpretation of such data is
still limited to LO pQCD, since NLO simulation programs are not yet
available.


\section{Quark Polarizations from Semi-Inclusive Measurements}
\begin{figure*} 
\begin{center}
\begin{picture}(300,215)
\put(-52,0){\epsfig{file=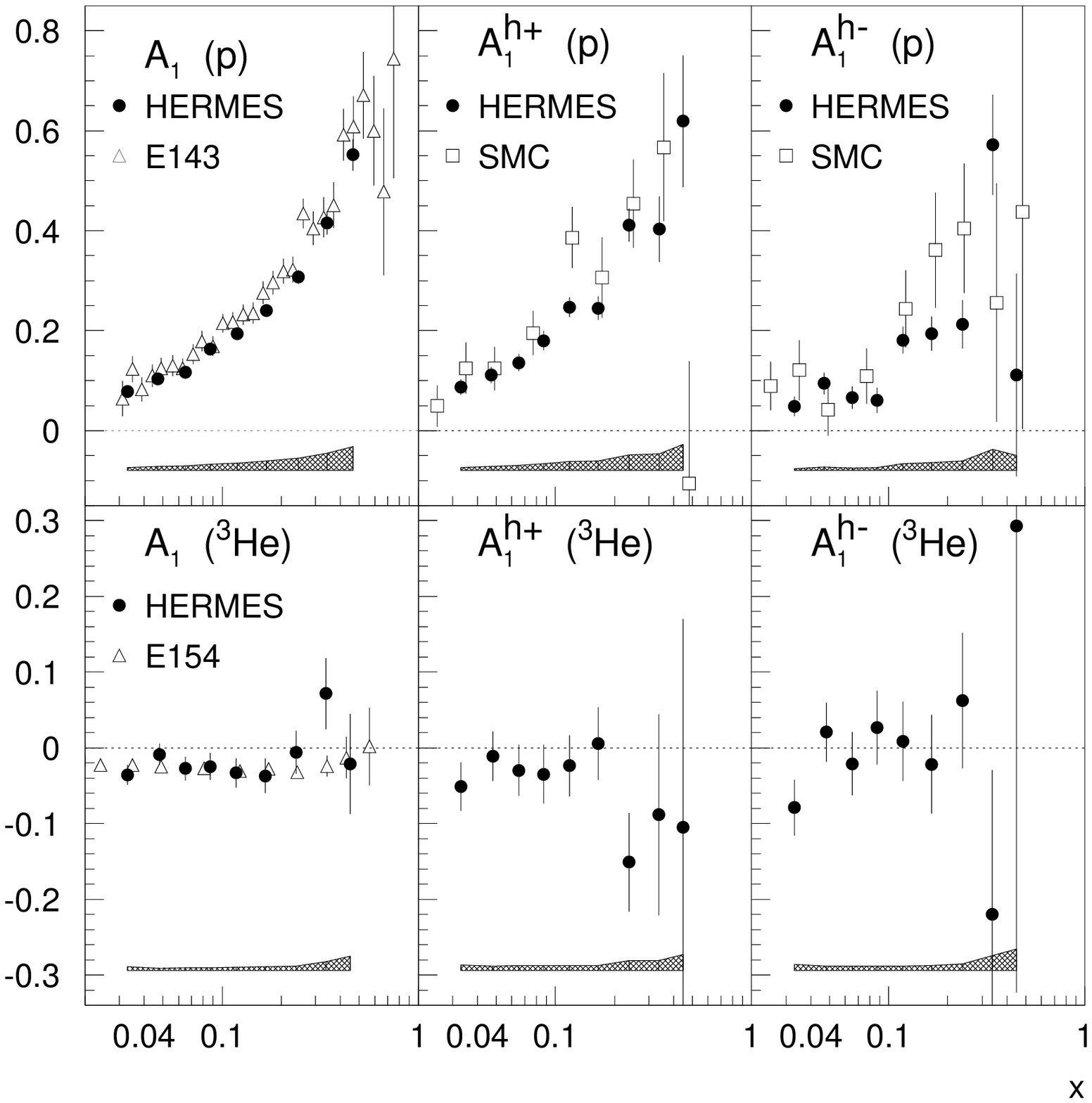,width=0.53\textwidth,bbllx=0pt,bblly=0pt,bburx=557pt,bbury=792pt}}
\put(145,-3.0){\epsfig{file=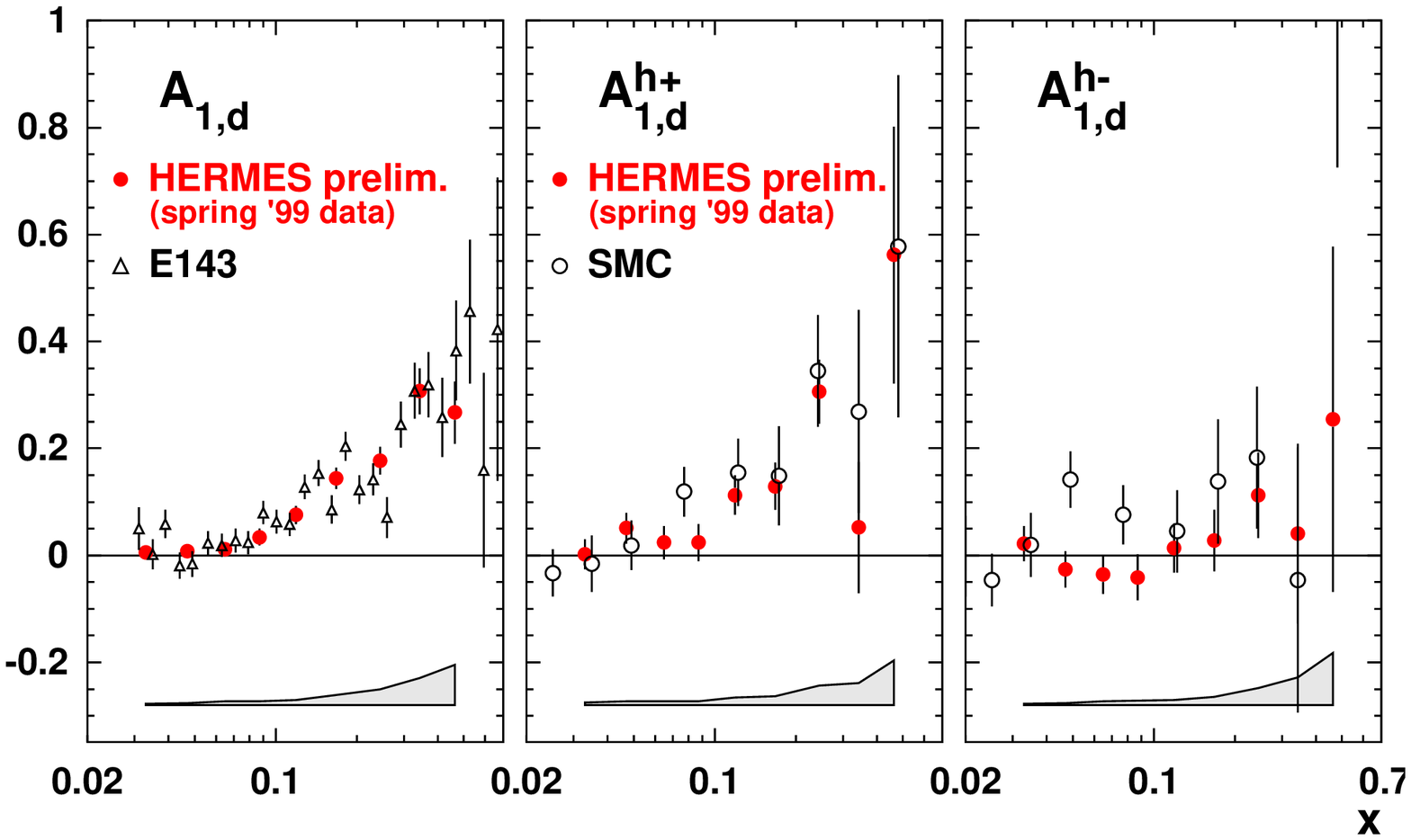,width=0.47\textwidth,bbllx=0pt,bblly=0pt,bburx=557pt,bbury=792pt}}
\put(-8,204.5){\footnotesize{(a)}}
\put(49,204.5){\footnotesize{(b)}}
\put(106,204.5){\footnotesize{(c)}}
\put(184,204.5){\footnotesize{(a)}}
\put(243,204.5){\footnotesize{(b)}}
\put(302,204.5){\footnotesize{(c)}}
\end{picture}
 \caption{ Inclusive (a) and semi-inclusive asymmetries for positively
 (b) and negatively (c) charged hadrons on the proton and $^3$He
 targets (left)~\protect\cite{us:Hermesdq} and on the deuterium target
 (right)~\protect\cite{us:Hermesasy}.  The bands represent the
 systematic uncertainties of the HERMES data.  \label{usfig:hermesasy}
 }
\end{center}
\end{figure*}

Information on the flavor separated polarized valence and sea quark
contributions can possibly be obtained via semi-inclusive scattering,
where one or more hadrons $h$ in coincidence with the scattered
charged lepton are detected.  According to the favored fragmentation
process, the charge of the hadron and its valence quark composition
provide sensitivity to the flavor of the struck quark as is
transparent within the QPM.  Hence the double-spin asymmetries for
hadrons, $A_1^h$, can be factored into separate $z$ and $x$ dependent
terms,
\bea \label{useq:a1h}
 A_1^h(x,Q^2) \approx   \displaystyle \frac{g_1^h}{F_1^h}(x,Q^2)= \hspace*{22mm}\\
 =\displaystyle  \frac{\int_{z_{min}}^1~dz\sum_q{e_q^2 {\Delta q(x,Q^2)}\cdot
{D_q^h(z,Q^2)} }} {\int_{z_{min}}^1~dz\sum_q{e_q^2 {q(x,Q^2)}\cdot
{D_q^h(z,Q^2)}}} , \nonumber
\eea      
where the fragmentation function $D_q^h(z,Q^2)$ is the probability
that the hadron $h$ originated from the struck quark flavor $q$. Here
$z = E_h/\nu$ is the hadron momentum fraction in the lab frame.

Results for semi-inclusive double-spin asymmetries measured with the
large forward spectrometers of  the  SMC~\cite{us:SMC_98b}
 and  the Hermes~\cite{us:Hermesdq,us:Hermesasy}  collaborations are shown in
Fig.~\ref{usfig:hermesasy} including new preliminary data on the
deuterium target. The inclusive data from Hermes in
Fig.~\ref{usfig:hermesasy}a are compared to SLAC (E143, E154) data.
The results agree well which reflects the understanding of the
experimental uncertainties.  All data points at a given $x$ have some
mean $ Q^2 $ which differs for the HERMES, SLAC and the SMC
experiments.  The comparison of the semi-inclusive data,
Figs.~\ref{usfig:hermesasy}b and c, thus points to a rather weak $Q^2$
dependence.  To maximize the sensitivity to the struck current quark,
typically kinematic cuts of $W^2 > 10$ GeV$^2$ and $z > 0.2$ are
imposed on the data in order to suppress effects from target
fragmentation.

\begin{figure*} 
\begin{center}
\includegraphics[width=\figsize]{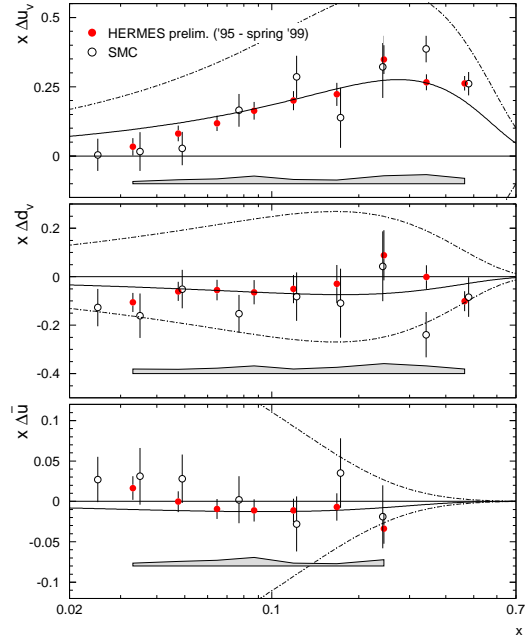}
 \caption{ Parton spin distributions at $Q^2=2.5$~GeV$^2$ for the
   valence quarks ${x\Delta u_v (x)}$, ${x\Delta d_v(x)}$ and the sea
   quarks $ {x\Delta \bar u(x)}$ as a function of $x$ from
   SMC~\protect\cite{us:SMC_98b} (data evolved to 2.5~GeV$^2$) and
   from Hermes with systematic uncertainty bands
   (preliminary)~\protect\cite{us:Hermesasy}.  The measurements are
   within the positivity limits given by the unpolarized quark
   distributions (dashed-dotted line) and are compared to
   parameterizations of the polarized quark distributions (solid
   line)~\protect\cite{us:GSgluon}.
 \label{usfig:hermesdq} }
\end{center}
\end{figure*}

According to Eq.~\ref{useq:a1h}, the double-spin asymmetries are
sensitive to the quark polarizations weighted with unpolarized
fragmentation functions.  Following the Hermes
analysis~\cite{us:Hermesdq}, a {\it flavor tagging} probability may be
determined by simulation using the LUND string fragmentation
model~\cite{us:lund}.  This allows in LO pQCD the polarized quark
distributions to be extracted, see Fig.~\ref{usfig:hermesdq}, using
the measured asymmetries from the various targets.  The present
asymmetry set is most sensitive to the light valence quark
polarizations ($\Delta {u_v}$, $\Delta {d_v}$) because of u(d)-quark
dominance: the production of $h^{\pm}$ is dominated by scattering off
u(d) quarks from a proton(neutron) target.  In particular, the impact
of new Hermes deuterium data on the precision of $\Delta {d_v}$ is
clearly seen in Fig.~\ref{usfig:hermesdq}.  However, the sensitivity
to the sea polarizations ($\Delta \bar{u}$, $\Delta \bar{d}$) is low,
less than 10\% at $x<0.2$.  Hence the sea polarization is assumed to
be flavor independent in present analyses.

Fig.~\ref{usfig:hermesdq}, as well as Fig.\ref{usfig:g1qcd1},
represent the first moments at a fixed $Q^2$ which determine the
flavor separated quark contributions to the nucleon spin: a positive
(parallel to the nucleon spin) up quark and a negative (anti-parallel
to the nucleon spin) down quark polarization are found.  The
polarization of the flavor undifferentiated sea is found to be
compatible with zero (see Fig.\ref{usfig:hermesdq}). It is
favored to be negative in spin-dependent NLO pQCD analyses (see
Fig.\ref{usfig:g1qcd1}). 

Theoretical conjecture (e.g. from the chiral
Quark-Soliton-Model~\cite{us:xqsm}) and recent attempts towards a
global analysis including semi-inclusive data (e.g. from
Ref.~\cite{us:GRSVsemi}) indicate a possible flavor asymmetry in the
nucleon's light sea, $\Delta \bar{u} - \Delta \bar{d} \ne 0 $, as in
the unpolarized case.  Current and future dedicated spin experiments
are expected to vastly broaden the information necessary for a
complete flavor separation.  Further improvement in the knowledge of
the sea polarizations will be soon available from the full high
statistics set of Hermes deuterium data employing the pion and kaon
identification capability of a RICH detector installed in 1998.


\section{Exclusive Spin Physics}
\begin{figure*} 
\begin{center}
\parbox{0.45\textwidth}
{ \begin{picture}(660,187)
\put(-100,-15){\epsfig{file=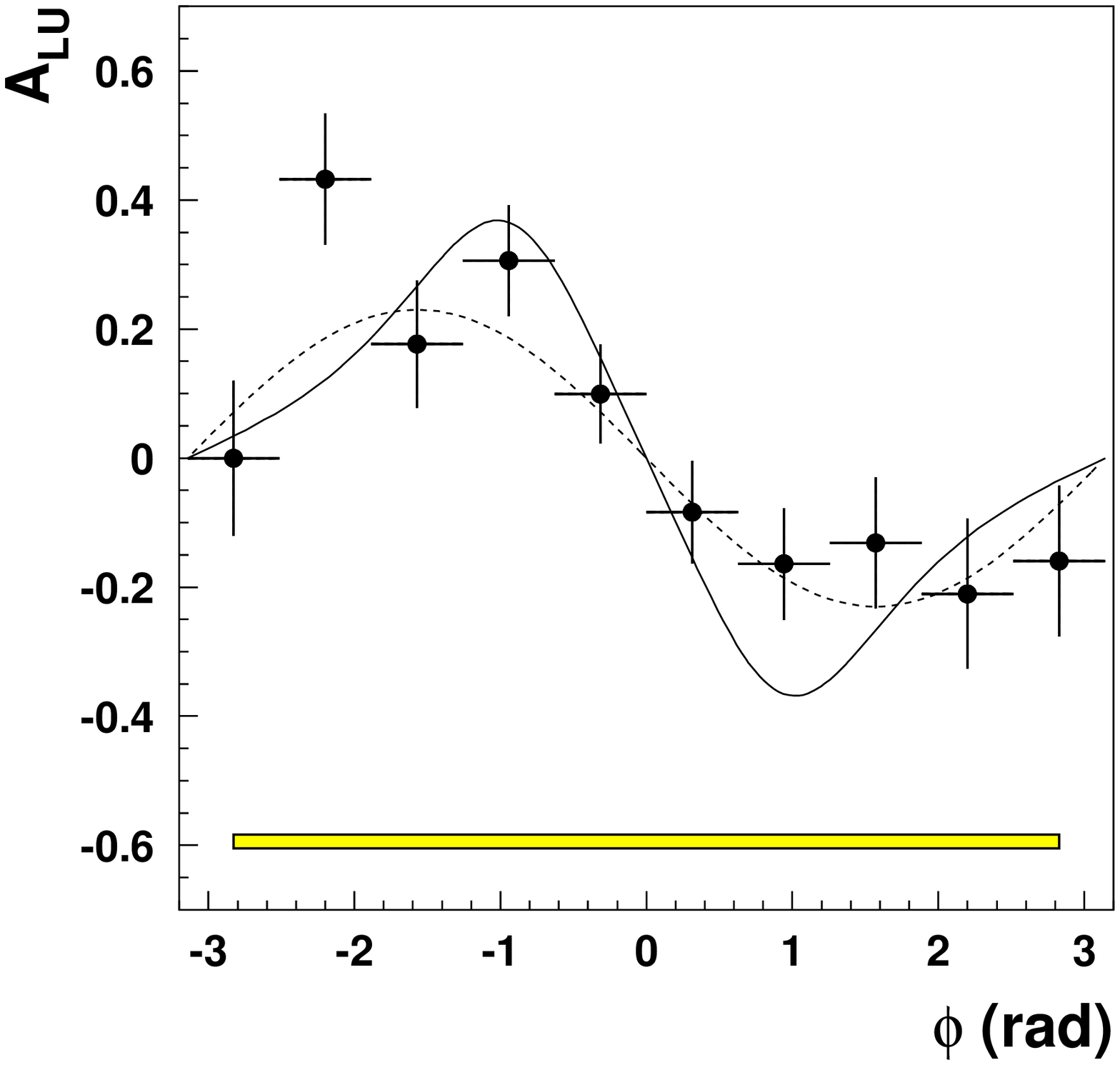,width=0.441\textwidth,bbllx=0pt,bblly=0pt,bburx=557pt,bbury=792pt}}
\put(108,12.5){\epsfig{file=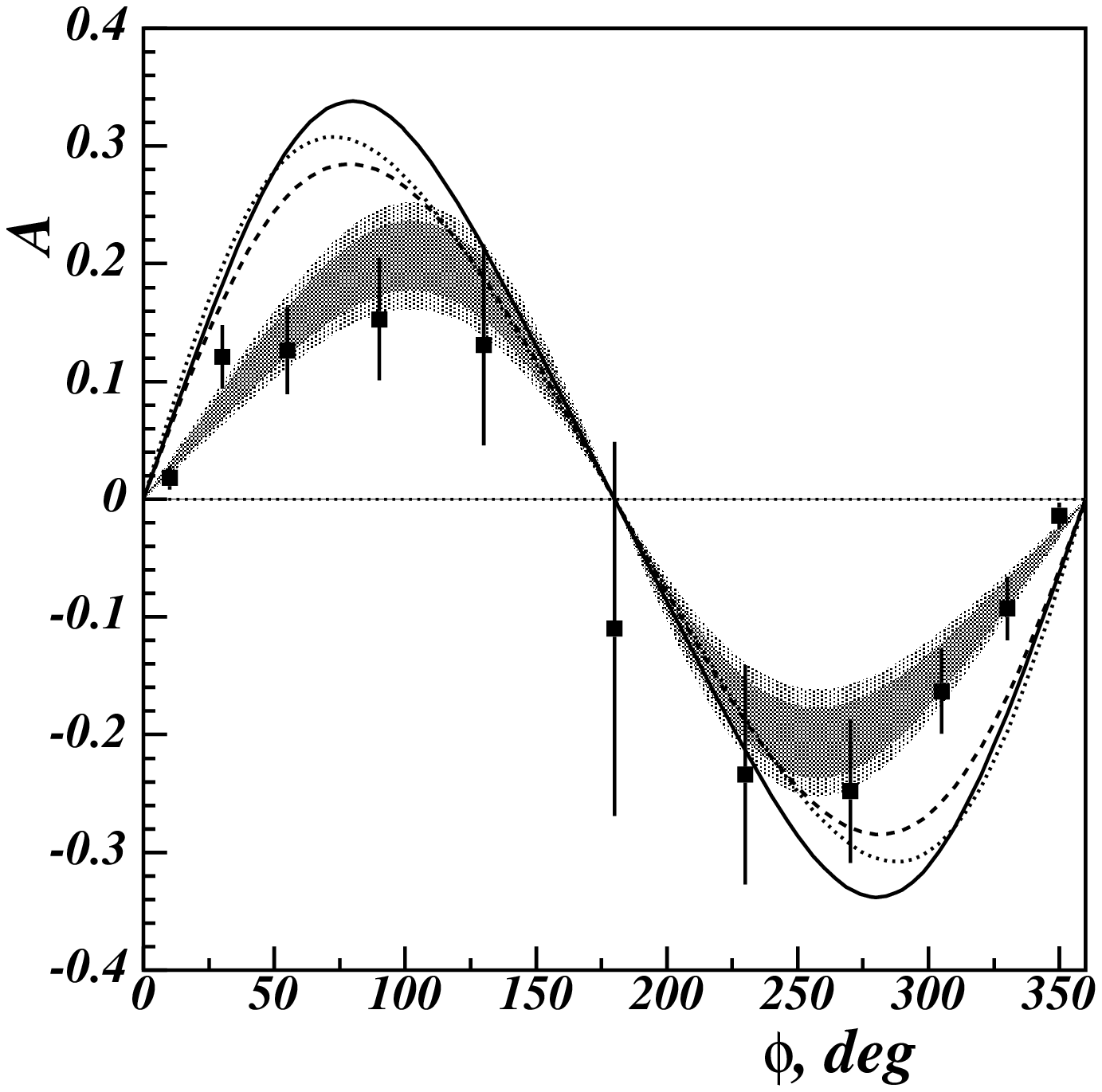,width=0.535\textwidth,bbllx=0pt,bblly=0pt,bburx=557pt,bbury=792pt}}
\end{picture}}
 \caption{ Single-spin asymmetry associated with DVCS as a function of
 $\phi$ from the Hermes~\protect\cite{us:dvcshermes} (left) and from
 the CLAS~\protect\cite{us:dvcsclas} (right) experiments. The curves
 represent model caluclations. The shaded area (right) represents the
 uncertainties of the best phenomenological $\phi$ dependent fit
 functions.  The subscripts L and U indicate the use of an
 longitudinally polarized beam and an unpolarized target,
 respectively. \label{usfig:dvcs} }
\end{center}
\end{figure*} 
Based on impressive theoretical efforts in the last decade, for the
first time a new window on the quark-gluon spin-structure of the
nucleon was opened with a description of hard exclusive processes as
deeply virtual Compton Scattering (DVCS) and meson production in QCD.
For a recent review see e.g. Ref.~\cite{us:exclreview}.  The framework
of DIS is here extendend to the non-forward region of the virtual
Compton process.  In the Bjorken limit the process may be viewed as
factorizing in two steps: a hard interaction of the virtual photon
with the nucleon, calculable in pQCD, and a soft interaction of the
struck quark with other partons containing new non-perturbative
information about the nucleon.  The non-perturbative functions are the
so-called generalized parton distribution functions.  GPDs (usually
denoted with $H,E, \tilde{H}, \tilde{E}$) represent probability
amplitudes to knock out a parton from a nucleon and to put it back
with a different longitudinal fraction of the momentum transfer $\xi$
(skewness parameter).  GPDs depend also on $Q^2$ and the squared
momentum transfer to the nucleon $t$.  These complex functions of
multiple variables unify known concepts of hadronic physics, e.g. by
linking ordinary parton distribution functions and nucleon form
factors. In the context of spin physics, the attractive fact is that
the total angular momentum $J_{q}$ and $J_{G}$ carried by a quark
flavor and by a gluon, respectively, are given by the second moment of
the sum of their unpolarized GPDs ($H^q,E^q$) in the limit
$t=0$~\cite{us:ji96}.  The total angular momenta of partons are still
unkown and related by angular momentum conservation to the spin of the
nucleon projected along an axis.  The latter could be written also as
a sum of contributions from quark ($ \Delta \Sigma$) and gluon
($\Delta G$) spin and orbital angular momentum of quarks ($L_{q}$) and
of gluons ($L_{g}$)~\cite{us:ji96,us:jaffe90}
\begin{equation}
\label{useq:SR}
 \frac{1}{2}=J_q + J_g = \frac{1}{2}\Delta\Sigma+L_q+\Delta G + L_g.
\end{equation} 
To understand the composition of the nucleon spin finally, each
contribution has to be identified.  The usefulness of such
angular momentum sum rule in the sense of a gauge invariant definition
and measurability of each of the terms are yet under
discussion~\cite{us:jaffe01}, especially with respect to the gluon.

The DVCS channel is viewed to be a particular clean hadronic reaction
that gives access to the GPDs~\cite{us:belitsky}.  In the case of hard
leptoproduction of mesons, however, the theoretical description
involves with the meson distribution amplitudes another unknown
non-perturbative input, which may complicate the identification of the
GPDs.  So far, the experimental data and GPD models focus on the
proton but GPDs for the deuteron were also introduced recently, see
e.g.  Ref~\cite{us:gpddeut}.

First experimental results on lepton beam helicity dependent
asymmetries associated with DVCS were obtained from Hermes and
confirmed by the CLAS collaboration, see Fig.~\ref{usfig:dvcs}.  This
single-spin asymmetry is sensitive to the interference term formed
from the imaginary part of the DVCS amplitude and the background QED
Compton (Bethe-Heitler) amplitude, $d\sigma_{\leftarrow}
-d\sigma_{\rightarrow} \propto {{ \mathrm{{Im}}
(\mathrm{{T}}_\mathrm{{DVCS}})\mathrm{T}_\mathrm{BH}}}$. 
In a LO leading twist description, this leads to a $ \sin \phi$
dependence in the related asymmetry, $A(\phi) = \alpha \,
\mathrm{sign(e)}\,\sin\phi$.  Here, $\mathrm{sign(e)}$ represents the
sign of the beam charge and $\phi$ is the angle between the $\gamma$
and e scattering planes.  The integrated beam-spin asymmetry obtained
from positron scattering off a hydrogen target at Hermes is 
${\alpha = -0.23 \pm 0.04\mathrm{ (stat)} \pm 0.03\mathrm{ (syst)} }$, 
while electron scattering at CLAS yields ${\alpha = 0.202 \pm
0.021\mathrm{ (stat)} \pm 0.009\mathrm{ (syst)}}$.  Both results are
in fair agreement with a simple $\sin\phi$ dependence and a change in
the sign is seen with the beam charge used.  The restricted kinematic
range do not allow to study dependencies on the relevant kinematic
variables so far. It is for Hermes, $\langle Q^2\rangle =
2.6$~GeV$^2$, $\langle \mathrm{Bjorken-}x \rangle = 0.11 $, and
$\langle -t\rangle = 0.27$~GeV$^2$, while for CLAS it is $1 $~GeV$^2 <
Q^2 < 1.75 $~GeV$^2$, $0.13 < \mathrm{Bjorken-}x < 0.35$, and
$0.1$~GeV$^2 < -t < 0.3 $~GeV$^2$. Both experiments will continue such
measurements and more precise data are expected to become
available. It will then particularly be interesting to test the
significance of higher $\sin \, \phi$ moments which are sensitive to
quark-gluon correlations as described by twist-three GPDs.
\begin{figure*}  
\begin{center} 
\begin{picture}(660,155)
\put(0,0) 
{{\epsfig{file=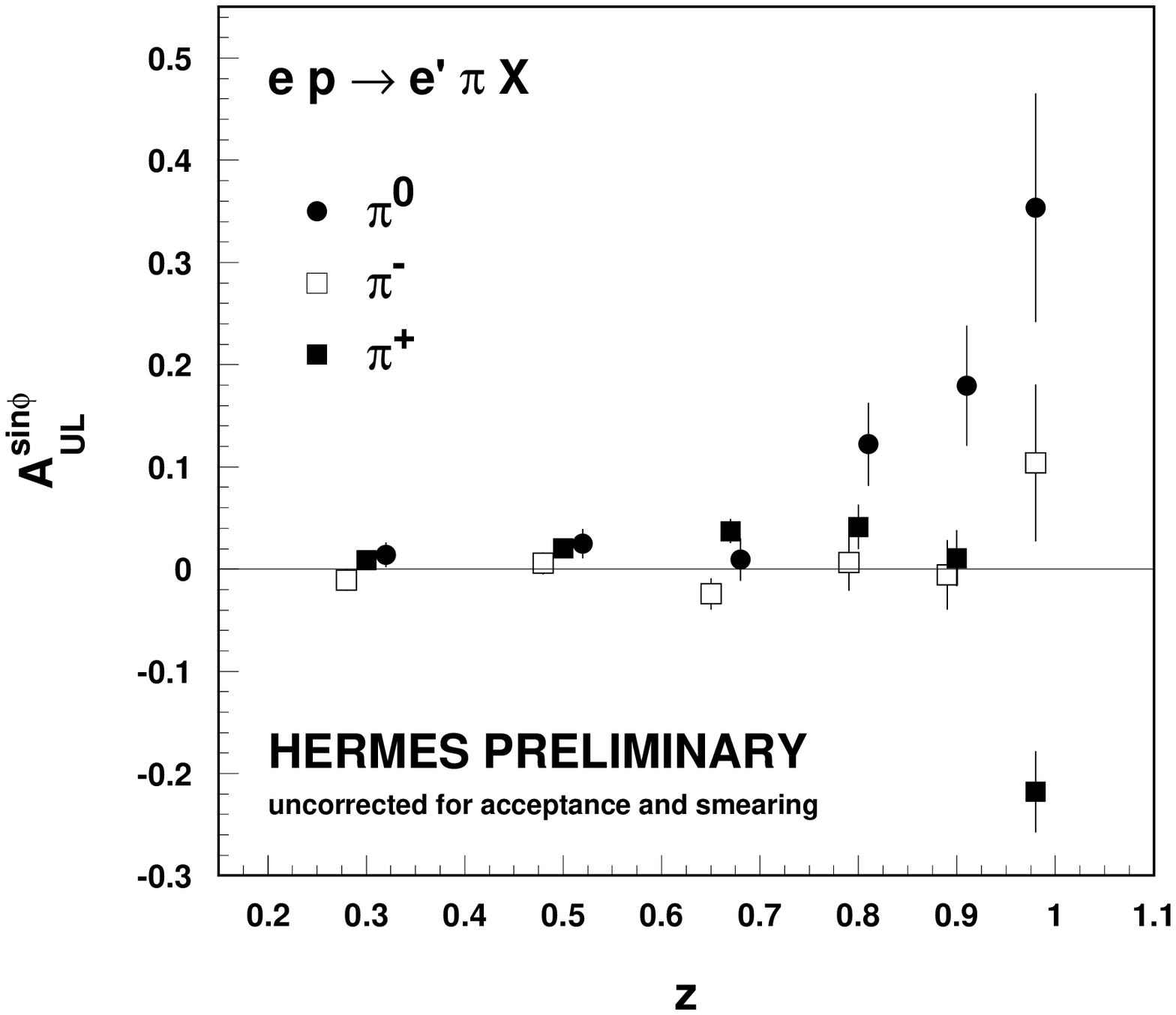,width=0.4433\textwidth}
\hspace*{1.3cm} 
{\put(0,-7){\epsfig{file=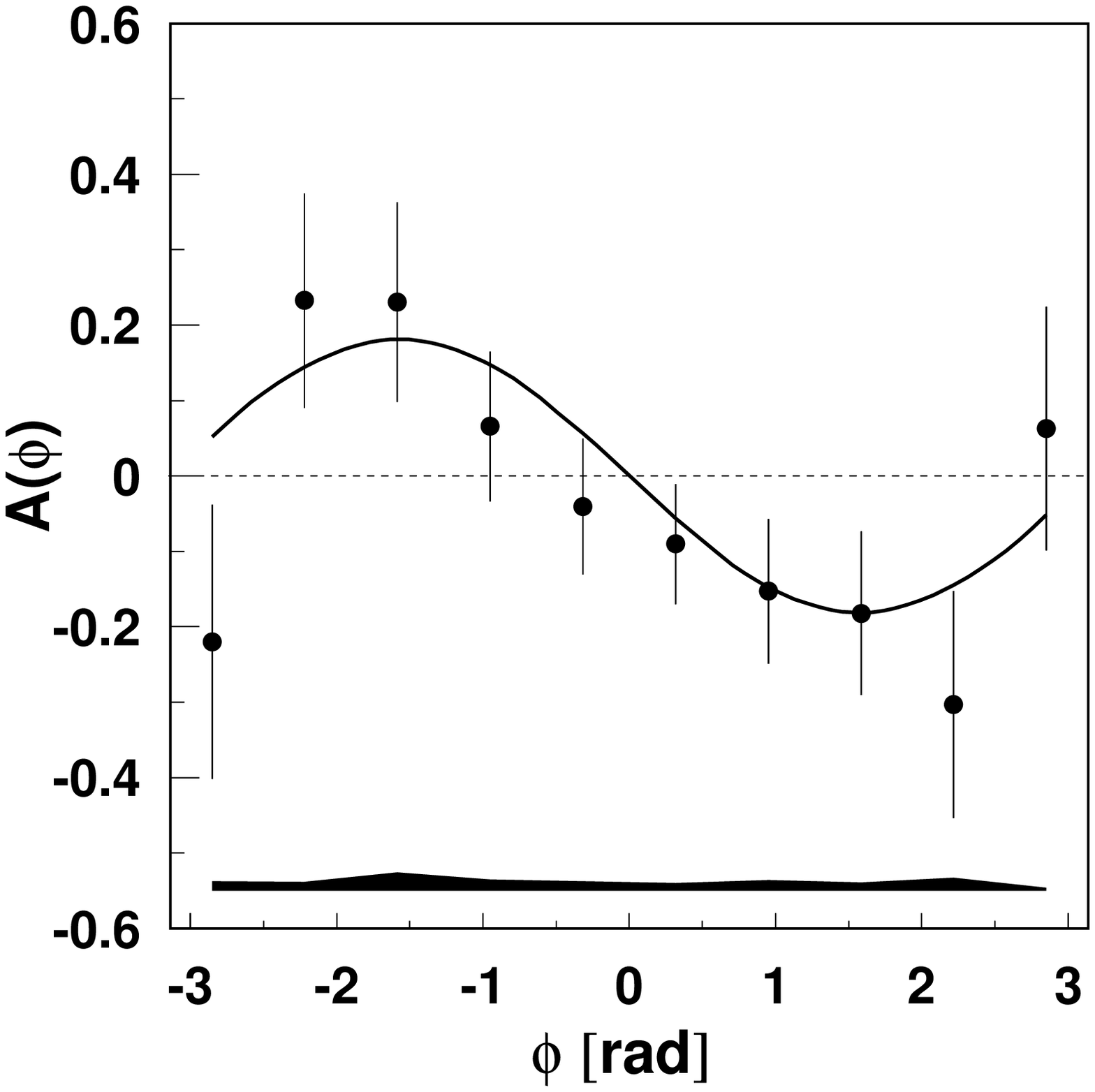,width=0.408\textwidth}}}
}} 
\end{picture}   
   \end{center} 
 \caption{ Hermes results  
   on azimuthal $\sin \phi$ moments of the single-spin asymmetries
   measured for the production of charged and neutral pions, as a
   function of the pion momentum fraction
   $z$~\protect\cite{us:avakian} (left) and for the exclusive
   $\mathbf{\pi}^+$ production~\protect\cite{us:thomas} (right).
   Exclusive $\pi^+$ was selected by requiring the missing mass $M_X$
   of the reaction $e^+ p \rightarrow e^{+\prime} \pi^+ X$
   corresponded to the nucleon mass, $M_X< 1.05$~GeV. The curve is a
   fit to the data by $A(\phi) = A_{\mathrm {UL}}^{\sin\phi} \cdot
   \sin\phi$. The subscripts U and L indicate the use of an
   unpolarized beam and a longitudinally polarized target,
   respectively.
 \label{usfig:pions} }
\end{figure*}  

Since only a quadratic combination of GPDs appears in the unpolarized
cross section, polarization is needed in order to disentangle the
various distributions by accessing additional observables.  For
example, it has been predicted~\cite{us:frankfurt} that for the
exclusive production of $\pi^+$ mesons from a transversely polarized
target by longitudinally virtual photons, the interference between the
pseudoscalar ($\tilde{E}$) and pseudovector ($\tilde{H}$ with
$\tilde{H} \rightarrow$ $\Delta q$ in forward limit at $t=0$)
amplitudes leads to a large asymmetry in the distribution of the angle
$\phi$.
Here $\phi$ is the azimuthal angle of the pion around the lepton
scattering plane.

For the first time, the Hermes collaboration presented data on
single-spin azimuthal asymmetries measured in the reaction $e p
\rightarrow e' \pi X$ using longitudinally polarized
protons~\cite{us:avakian,us:thomas}, see Fig.~\ref{usfig:pions}.
Comparing the left and right panels of Fig.~\ref{usfig:pions} it is
clearly seen, that for $\pi^+$ production at high $z$ ($>0.9$) the
$\sin \phi$ moment of the asymmetry is observed~\cite{us:avakian} to
become suddenly negative, in agreement with a different (but related)
analysis of exclusive $\pi^+$ data~\cite{us:thomas}.  A fit to the
exclusive data, $A(\phi) = A_{\mathrm {UL}}^{\sin\phi} \cdot
\sin\phi$, delivers the $\sin \phi$ moment of the target-spin related
asymmetry to be $A_{\mathrm {UL}}^{\sin\phi} = -0.18 \pm 0.05
$(stat)$\pm 0.02 $(syst) integrated over the kinematic range, i.e.
$\langle Q^2\rangle = 2.2$~GeV$^2$, $\langle \mathrm{Bjorken-}x
\rangle = 0.15 $, and $\langle -t\rangle = 0.4 $~GeV$^2$.

In the case of electroproduction from a target polarized
longitudinally with respect to the lepton beam momentum, a small (at
Hermes about 20\%) transverse component ($S_{\perp}$) of the target
polarization with respect to the direction of the virtual photon is
present along with the dominating longitudinal component ($S_{||}$).
According to Ref.~\cite{us:vanderh}, $A_{\mathrm {UL}}^{\sin\phi}$
occurs in the polarized cross section $\sigma_S$ of the reaction
\begin{equation}
\label{useq:azipions}
\sigma_S \sim [ S_{\perp} \sigma_{\cal{L}} + S_{||}\sigma_{\cal{LT}} ] A_{\mathrm {UL}}^{\sin\phi} \, \sin\phi, 
\end{equation} 
where contributions arise from the longitudinal ($\cal{L}$) amplitude
and from the interference ($\cal{LT}$) of longitudinal and transverse
($\cal{T}$) photon amplitudes, the latter being suppressed by $1/Q$
relative to $\sigma_{\cal{L}}$. However, both terms in
Eq.~\ref{useq:azipions} are expected to contribute at the same order
in $1/Q$ since $\sigma_{\cal{L}}$ is weighted with the $1/Q$
suppressed transverse spin component $S_\perp$.  For a complete
interpretation of the measured asymmetry quantitative predictions for
the term $\sigma_{\cal{LT}}$ are required which rely on
next-to-leading twist calculations~\cite{us:mueller}. The upcoming
Hermes data on a transversely polarized hydrogen
target~\cite{us:vincter} are important for further theoretical
understanding and a decomposition of the contributions from the two
target spin components.

\section{Transversity and Prospects of Current Experiments } 
\label{us:secprospects}

Regarding new experiments one of the most interesting questions is to
determine the still unknown third type of twist-two quark distribution
function $\delta q$, called transversity, first mentioned in
Ref.~\cite{us:ralston}. In order to probe the transverse spin
polarization of the nucleon, a helicity (identical to chirality at
leading twist) flip of the struck quark must have occured.  In hard
processes this is only possible with non-zero quark masses, thus
suppressing this function in inclusive deep inelastic scattering.
However, in semi-inclusive processes it is possible to combine two
chiral odd parts, one describing the quark content of the target
($\delta q$) and another one describing the quark fragmentation into
hadrons.  Considerable effort has gone into understanding, modelling
and proposed measures of $\delta q$, for a review see e.g.
Ref.~\cite{us:transreview}.

There are important differences to be noted between the helicity and
the transversity distributions which give further insight into the
non-perturbative QCD regime of hadronic physics. For example, as
mentioned before, quark and gluon helicities mix under $Q^2$
evolution, but there is no analog of gluon transversity in the
nucleon.  Furthermore, the difference between $\Delta q$ and $\delta
q$ reflects the relativistic character of quark motion in the nucleon.
Only in the case of non-relativistic movement of quarks in the nucleon
are $\Delta q$ and $\delta q$ identical, i.e. invariant under a series
of boosts and rotations which convert the longitudinally polarized
nucleon into a transversely one.  The first moments of the
transversity distributions for quarks and anti-quarks are related to
the flavor dependent contribution to the nucleon tensor charge $\delta
\Sigma$, which behaves as a non-singlet matrix
element~\cite{us:jaffeji}.  Hence the tensor charge is expected to be
a more quark-model-like quantity in contrast to the axial charge, but
more difficult to predict~\cite{us:gamberg}.

\begin{figure} 
\begin{center}
\includegraphics[width=\figsizeb]{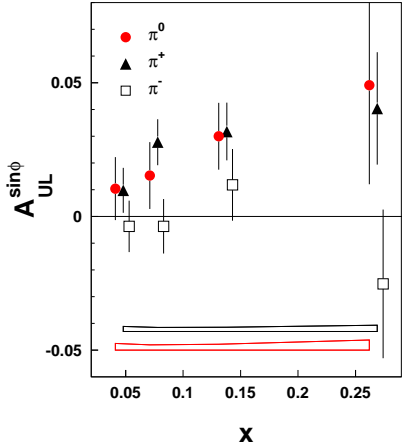}
\caption{ 
  Analyzing power in the $\sin \phi$ moment for Hermes data on
  semi-inclusive charged~\protect\cite{us:Herm00} (systematic
  uncertainty: lower band) and neutral~\protect\cite{us:Herm01}
  (systematic uncertainty: upper band) pion production on a
  longitudinally polarized hydrogen target.
 \label{usfig:semipions} }
\end{center}
\end{figure}

The recent observation of non-zero single-spin azimuthal asymmetries
for neutral and positively charged pions by the Hermes collaboration
generated much interest, since it can be interpreted as evidence for a
non-zero chirality-flipping fragmentation function that couples to the
quark transversity distributions.  The
results~\cite{us:Herm00,us:Herm01} are presented in
Fig.\ref{usfig:semipions}, see also the left panel of
Fig.\ref{usfig:pions} and $z<0.7$.  The data were taken with a
longitudinally polarized target which makes the interpretation of the
Hermes results difficult due to possible additional twist-three
contributions.

The study of transversity distributions and chiral-odd fragmentation
functions, at least for the up-quark with good precision, is a primary
goal of the Hermes Run II~\cite{us:vincter} using a transversely
polarized hydrogen target. Transversity is also an important part of
ongoing and forthcoming experiments, see Ref.~\cite {us:transzeuthen}
for an overview of the experimental state of the field.  At BNL-RHIC
interesting processes involving transversity in pp collisions are
Drell-Yan lepton pair production with two protons transversely
polarized or alternatively, chiral-odd two-pion interference
fragmentation in large $p_T$ pion pair production using one proton
transversely polarized.

The COMPASS~\cite{us:heinsius} experiment at CERN has a transversity
programme similiar to that of Hermes covering a different kinematic
region. The primary goal of the COMPASS muon programme is the
measurement of the gluon polarization $\Delta G/G$ with $\sim 0.1$
accuracy via open charm production and hadron pair production at large
$p_T$ for $0.04 < x_\mathrm{G} < 0.3$ about.

Using both colliding proton beams longitudinally polarized, prompt
photon production will be employed at RHIC to measure the
helicity-dependent gluon density $\Delta G$, at $0.02 < x_\mathrm{G} <
0.3$~\cite{us:bland}.  A compilation of simulated statistical
accuracies for $\Delta G/G$ may be found e.g. in
Ref.~\cite{us:future}.  It is important that various channels for
extracting $\Delta G$ in eN- and pp-scattering are required to
minimize the (so far strong) model-dependencies.

Alternatively to the Hermes programme to measure the flavor separated
quark distributions, the production of weak $W^\pm$ bosons in high
energy polarized pp collisions at RHIC provides sensitivity to the
quark and antiquark spin distributions~\cite{us:vigdor}. The maximal
parity violation in the interaction and the dependence of the
production on the weak charge of the quarks may be used to select the
specific flavor and charge of the quarks.

\section{Concluding Remarks}

Spin physics remains an exciting, rapidly developing field of research
and contributes remarkably to the QCD picture of the structure of the
nucleon.  Recent precise spin structure function data from DESY and
SLAC together with previous data improve the knowledge about the
contribution of valence quarks to the nucleon spin within the
framework of NLO pQCD and allow the fundamental Bjorken sum rule to be
tested. Newest semi-inclusive double-spin asymmetry data from Hermes
deliver additional sensitivity required for a complete flavor
separation of polarized parton distributions, so far in LO pQCD.  The
contribution of gluons to the nucleon spin is not yet well known.  It
is suggested to be positive from LO/NLO pQCD fits based on inclusive
DIS data and from a LO pQCD interpretation of Hermes high $p_T$ hadron
pair production data.  A recent interesting development in QCD spin
physics was triggered by the Hermes measurement of single-spin
azimuthal asymmetries in semi-inclusive pion electroproduction off a
longitudinally polarized target.  This observation suggests a non-zero
chiral-odd fragmentation function which allows one to access the so
far unknown quark transversity distribution in semi-inclusive
scattering from transversely polarized targets.  For the first time, a
window may be opened to access angular momenta of partons using the
framework of generalized parton distribution functions based on new
data on spin-dependent, hard exclusive processes, released by the
Hermes and CLAS collaborations.  Further experimental studies of the
connection of semi-inclusive with exclusive reactions, and of high
energy with low energy spin physics are being performed at JLAB and at
DESY.

More precise data are expected to soon become available on the gluon
spin, on the flavor separated quark and anti-quark helicities and on
transversity properties from the high luminosity experiments at CERN,
DESY, JLAB and RHIC-Spin, and also from an upcoming SLAC
experiment~\cite{us:slac-161}.  The perspectives also of future
polarized lepton-nucleon fixed-target~\cite{us:future,us:ryckbosch}
and collider~\cite{us:eic} experiments are being discussed
intensively.  The goal remains to be the development of a complete,
firm theoretically picture of the momentum and spin structure of
hadrons.

\section*{Acknowledgment}
I would like to thank D. Ryckbosch and E. Kinney for carefully reading
this manuscript.  Many thanks to Juliet Lee-Franzini and her wonderful
team for organizing such an inspiring conference.


\end{document}